\documentclass[%
 reprint,
 superscriptaddress,
 amsmath,amssymb,
 aps, floatfix
]{revtex4-1}

\usepackage{comment}
\usepackage{graphicx}
\usepackage{multirow}
\usepackage{dcolumn}
\usepackage{bm}
\usepackage{xcolor}
\usepackage[version=4]{mhchem}

\newcommand{\pcond}[2]{\left({#1}\middle|{#2}\right)}

\newcommand{\ba}{\mathbf{a}}

\newcommand{\by}{\mathbf{y}}

\newcommand{\ang}{\textup{\AA}}

\begin{document}

\title{A Bayesian approach to NMR crystal structure determination}

\author{Edgar A. Engel}
 \affiliation{Laboratory of Computational Science and Modeling, Institut des Mat\'eriaux, \'Ecole Polytechnique F\'ed\'erale de Lausanne, 1015 Lausanne, Switzerland}
\author{Andrea Anelli}
 \affiliation{Laboratory of Computational Science and Modeling, Institut des Mat\'eriaux, \'Ecole Polytechnique F\'ed\'erale de Lausanne, 1015 Lausanne, Switzerland}
\author{Albert Hofstetter}
 \affiliation{Laboratory of Magnetic Resonance, Institut des Sciences et Ing\'enierie Chimiques, \'Ecole Polytechnique F\'ed\'erale de Lausanne, 1015 Lausanne, Switzerland}
\author{Federico Paruzzo}
 \affiliation{Laboratory of Magnetic Resonance, Institut des Sciences et Ing\'enierie Chimiques, \'Ecole Polytechnique F\'ed\'erale de Lausanne, 1015 Lausanne, Switzerland}
\author{Lyndon Emsley}
 \email{lyndon.emsley@epfl.ch}
 \affiliation{Laboratory of Magnetic Resonance, Institut des Sciences et Ing\'enierie Chimiques, \'Ecole Polytechnique F\'ed\'erale de Lausanne, 1015 Lausanne, Switzerland}
\author{Michele Ceriotti}
\email{michele.ceriotti@epfl.ch}
 \affiliation{Laboratory of Computational Science and Modeling, Institut des Mat\'eriaux, \'Ecole Polytechnique F\'ed\'erale de Lausanne, 1015 Lausanne, Switzerland}

\date{\today}
\begin{abstract}

Nuclear Magnetic Resonance (NMR) spectroscopy is 
particularly well-suited to determine the structure of molecules and materials in powdered form. 
Structure determination usually proceeds by finding the best match between experimentally observed NMR chemical shifts and those of candidate structures.
Chemical shifts for the candidate configurations have traditionally been computed by electronic-structure methods, and more recently predicted by machine learning. 
However, the reliability of the determination depends on the errors in the predicted shifts. Here we propose a Bayesian framework for determining the confidence in the identification of the experimental crystal structure, based on knowledge of the typical error in the electronic structure methods. We also extend the recently-developed ShiftML machine-learning model, including the evaluation of the uncertainty of its predictions.
We demonstrate the approach on the determination of the structures of six organic molecular crystals. We critically assess the reliability of the structure determinations, facilitated by the introduction of a visualization of the of similarity between candidate configurations in terms of their chemical shifts and their structures. 
We also show that the commonly used values for the errors in calculated $^{13}$C shifts are underestimated, and that more accurate, self-consistently determined uncertainties make it possible to use $^{13}$C shifts to improve the accuracy of structure determinations. 
\end{abstract}

\maketitle

\section{\label{sec:intro}Introduction}

Solid-state nuclear magnetic resonance (NMR) crystallography is a powerful tool for determining the atomic-level structure and dynamics of solids. Since the values of chemical shifts measured for different nuclei directly reflect the corresponding local environments, it does not rely on the presence of long-range order, rendering it especially suitable for powdered and amorphous solids. 
This has led to widespread application in many fields ranging from materials science to pharmaceutical chemistry~\cite{salager_2010_nmr, baias_2013_nmr, baias_2013_csp, kalakewich_2013, brouwer_2013, ashbrook_2008, kevern_2009, dedios_1993, harper_2006, witter_2002, ochsenfeld_2001_nmr, widdifield_2016_nmr, hartman_2016, castellani_2002, mollica_2015, santos_2013_nmr, brunet_2004, widdifield_2005, lai_2011, farnan_1992, seymour_2016, brouwer_2008, cadars_2014}.

In 
\textit{de novo} structure determinations based on solid-state NMR data, the most powerful approach today is to compare experimental data and calculated NMR chemical shifts to identify which structure among an ensemble of candidates corresponds to the experimental sample~\cite{baias_2013_csp,baias_2013_nmr,harper_2006,harper_2006_nmr,widdifield_2016_nmr,widdifield_2017_nmr,mueller_2013}.
The pool of candidates can be generated either by comprehensive crystal structure prediction (CSP)~\cite{baias_2013_csp, baias_2013_nmr, widdifield_2017_nmr, hofstetter_2019_nmr}, or through searches incorporating different degrees of intuition, or complementary experimental data~\cite{kumar_2017, mueller_2013, harper_2006, harper_2003, baias_2015, leclaire_2016_nmr}. 
Irrespective of the exact method, there are usually at least dozens of plausible structures.
For these candidates NMR shifts are calculated, most often using gauge-including projector-augmented wave (GIPAW) DFT~\cite{blochl_1994_gipaw,pickard_2001_gipaw,yates_2007_gipaw} calculations or, more recently, machine-learning (ML) models trained on GIPAW reference data~\cite{paruzzo_2018_shiftml}. 
The accuracy of these methods in reproducing the subtle dependencies of chemical shifts on differences in local atomic environments underlies the widespread success of chemical shift based NMR crystal structure determinations~\cite{ochsenfeld_2001_nmr, harris_2004_nmr, harper_2006_nmr, harris_2007_nmr, othman_2007_nmr, salager_2009_nmr, salager_2010_nmr, webber_2010_nmr, dudenko_2012_nmr, baias_2013_nmr, pawlak_2013_nmr, santos_2013_nmr, ludeker_2015_nmr, paluch_2015_nmr, watts_2016_nmr, widdifield_2016_nmr, mali_2017_nmr, harris_2006_nmr, mifsud_2006_nmr, heider_2007_nmr, baias_2013_nmr, fernandes_2015_nmr, leclaire_2016_nmr, selent_2017_nmr, widdifield_2017_nmr, nilsson_2018_nmr, rawal_2010, charpentier_2011, jiang_2011, bonhomme_2012, baias_2013_csp, kalakewich_2013, mueller_2013, ashbrook_2016}. 

While usually sufficiently accurate, GIPAW shifts are not exact and the underlying atomic structures of candidates is subject to the accuracy of the level of theory at which they are described, leading to uncertainties in predicted NMR shifts~\cite{salager_2010_nmr}.
Conventionally a structure is therefore considered to be consistent with experiment if the root-mean-square deviation (RMSD) of its shifts from the experimentally measured values falls within these uncertainties.
However, this approach is severely limited. It neither allows determination of the experimental structure when multiple candidates exhibit similar RMSDs within the ``confidence interval'', nor does it provide a means of quantifying how likely different candidates are to match the experimental structure in any but the most clear-cut cases.
To overcome these limitations here we propose a probabilistic approach to the evaluation of candidate structures in NMR crystal structure determination.
Whereas the conventional approach simply determines whether structures agree with the data or not, the new method allows one to quantitatively evaluate the probability that a structure among a given set corresponds to the experiment, on a continuous scale from 0 to 100\% confidence. 
As a demonstration of the capabilities of the method, we combine experimental NMR data with GIPAW and ML predictions of the shifts of a set of CSP candidates to determine the confidence in the structure determination of six different molecular crystals. 
We find that the structures of ampicillin, flufenamic acid, cocaine, and AZD8329 can be identified with very high confidence (between 91\% and 100\%). In contrast, we show that the determination of the structure of flutamide is less certain (82\% confidence) and confirm the low confidence (13\%) in the capability to determine the structure of theophylline~\cite{baias_2013_csp}.
We further introduce a method to visualise the Bayesian probabilities of the candidate structures in combination with a low-dimensional representation of their similarity, computed according to their chemical shifts or their geometry. 
A critical analysis of the impact of different details of the NMR crystallography protocol on the reliability of the structural determination allows us to confirm the importance of pre-determining the assignment of shifts to specific molecular moieties in a structure and of assessing self-consistently the uncertainty in the DFT (or ML) shifts.
In particular, we find that for the compounds considered here the errors in the calculated $^{13}$C shifts are substantially larger than literature estimates of the uncertainty in $^{13}$C shifts, and that with self-consistently determined uncertainties the inclusion of $^{13}$C shifts (in addition to $^1$H shifts) leads to more reliable structure determinations.

\section{\label{sec:theory}Theory}

In our 
probabilistic approach to NMR crystal structure determination each candidate structure constitutes a ``model'', $M$, for which we determine the posterior probability, $p(M|{\bf y}^\star)$, of corresponding to the experimental structure, given experimentally determined shifts, ${\bf y}^\star$.
The experimental shifts may originate from a single or multiple chemical species and may or may not have been partially or fully assigned to particular nuclei within the compound of interest.
For each model the prior probability of matching the experimental structure is denoted by $p(M)$ and can in principle incorporate information regarding the thermodynamic stability of different candidates.
Noting that stability estimates are often not fully trustworthy on the scale of differences between models, here we choose to set aside such considerations and assume uniform priors for all $n_M$ models, $p(M)=1/n_M$.

We denote the probability of observing shifts ${\bf y}$ for a given model $M$ as $p({\bf y}|M)$
and the probability of observing a shift ${\bf y }$ \textit{before} we run the experiment as $p({\bf y})=\sum_{M} p({\bf y}|M) p(M)$.
Bayes theorem dictates that 
\begin{equation}
p(M|{\bf y}^\star) = \frac{p({\bf y}^\star|M)p(M)}{p({\bf y}^\star)} =  \frac{p({\bf y}^\star|M)p(M)}{\sum_{M'} p({\bf y}^\star|M')p(M') } \quad.
\label{eq:posterior-base}
\end{equation}
Clearly, in order to assess %
evaluate the posterior $p(M|{\bf y}^\star)$, 
the conditional probability distribution $p({\bf y}|M)$ must be defined. 
Given GIPAW or ML estimates of the shifts ${\bf y}^M$ for each model $M$, the simplest model for the conditional distribution of the shift associated with a particular nucleus $j$ takes the form of a normal distribution
\begin{equation}
\label{eq:gaussian-pjym}
p_j(y|M) = \frac{1}{\sqrt{2 \pi \sigma_j^2}} \exp{\left(-\frac{1}{2}\left(\frac{y - y^M_j}{\sigma_j}\right)^2\right)} \quad.
\end{equation}
The width $\sigma_j$ represents an estimate of the typical error in the calculated shift with respect to experiment. 
We will discuss different approaches to determining $\sigma_j$ later, and will start by discussing how to translate Eq.~\eqref{eq:gaussian-pjym} into 
a posterior $p(M|{\bf y}^\star)$, which quantifies the confidence in designating the model $M$ as the experimental structure.
\subsection{\label{subsec:fullassign}With full assignment of shifts}
In order to evaluate $p(M|{\bf y}^\star)$, one needs to combine information from all experimental shifts $\by^\star = \{ y_j^\star \}$, determining the conditional probability $p\pcond{\by^\star}{M}$ based on the probabilities for individual shifts in Eq.~\eqref{eq:gaussian-pjym}.
In the simplest case a full assignment of the experimental shifts to the nuclei in the compound has been determined, for example through methods such as those described in Ref.~\cite{baias_2013_nmr,hofstetter_2019_nmr}.
Assuming independent errors on shifts from distinct nuclei, $p\pcond{\by^\star}{M}$ becomes
\begin{equation}
 p\pcond{\by^\star}{M} = \prod_j p_j\pcond{y^\star_j}{M}.
\end{equation}

\subsection{\label{subsec:noassign}Without assignment of shifts}
Although the default scenario will involve full assignments of experimental shifts to particular nuclei, in rare cases definitive assignments may not be available.
One must then consider the different ways of assigning the experimental shifts. If the permutation vector that describes one such assignment is denoted as $\ba$, the conditional probability may be written as a sum over assignments,
\begin{equation}
p\pcond{\by^\star}{M}  = \sum_{\ba} p\pcond{\by^\star}{M, \ba} p\pcond{\ba}{M},
\label{eq:all-matches}
\end{equation}
where one can define the  conditional probability for a given assignment as 
\begin{equation}
p\pcond{\by}{M, \ba} = \prod_j p_{a_j}\pcond{y_j}{M}.
\end{equation}
If there is no heuristic way to determine the likelihood of a given assignment, $p\pcond{\ba}{M}$ has to be set to a constant. 
In this case, if one defines the matrix of conditional probabilities $P_{ij}=p_i\pcond{y_j}{M}$, $p\pcond{\by^\star}{M}$ is proportional to the permanent of the matrix, $p\pcond{\by^\star}{M} = \operatorname{perm}{\mathbf{P}}/n!$.

\subsection{\label{subsec:partial}Partial assignments}

Cases in which 
none of the experimental shifts can be assigned are rare.
In most cases the sum in Eq.~\eqref{eq:all-matches} only needs to be evaluated over a subset of all the possible permutations of indices $\ba$. 
In practice this means that $\mathbf{P}$ can be made block-diagonal, each block $\mathbf{P}_k$ corresponding to a group of nuclei that are distinct from the rest, but for which assignments among them are not available. 
The overall conditional probability can be written as a product between the permanents of the blocks 
\begin{equation}
p\pcond{\by^\star}{M}  = \prod_{k} \operatorname{perm}{\mathbf{P}_k}/n_k!,
\label{eq:partial-matches}
\end{equation}
where $n_k$ indicates the size of the $k$-th block.

While evaluating the permanent has a cost that grows combinatorially with the size of $\by^\star$, algorithms with a low prefactor make its evaluation affordable up to a few tens of nuclei (per block $k$). 
In extraordinary cases where its evaluation is not possible, a pragmatic but generally inaccurate alternative is to assume Eqs.~\eqref{eq:all-matches} and \eqref{eq:partial-matches} to be dominated by the contribution from the assignment producing the best-match between $\by^M$ and $\by^\star$.

\subsection{\label{subsec:gipaw_errors}Estimate of the reference errors}

Clearly, the evaluation of $p\pcond{M}{\by^\star}$ requires an estimate of the uncertainties $\sigma_j$ in calculated shifts.
Assuming that any errors in the experimental determination of the shifts can be neglected, there are still multiple sources of errors to consider. 
First, experimental shifts average over thermal and quantum fluctuations, while GIPAW shifts are usually calculated for the nearest local energetic minimum. 
Second, approximations in the description of the electronic structure lead to errors in the 
predicted shifts.
Third, errors are incurred by the conversion of the chemical shieldings obtained from GIPAW calculations (and ML models trained thereon) into chemical shifts via on-the-fly linear regression or a linear mapping driven by reference data~\cite{salager_2010_nmr,hartman_2016}. 
Finally, when using a ML model, an environment-dependent statistical error relative to the GIPAW reference is added on top of the underlying theory/experiment discrepancy. 

The statistical error, $\sigma_j^\text{ML}$, can be characterised efficiently and accurately (see appendix~\ref{app:mlmodel}), but estimating the error of the underlying GIPAW shifts with respect to experiment, $\sigma_j^\text{DFT}$, usually requires extensive benchmarks. Existing datasets~\cite{salager_2010_nmr,hartman_2016,dracinsky_2019} suggest that the typical errors are of the order of $\sigma_{\ce{H}}^\text{DFT} = 0.33 \pm 0.16$\,ppm, and  $\sigma_{\ce{C}}^\text{DFT} = 1.9 \pm 0.4$\,ppm.
As an alternative to these estimates, one can assess $\sigma_j$ for a specific molecule by considering $p_j\pcond{y}{M}$ to depend parametrically on the uncertainty $\sigma_j$ and maximizing $p(\by^\star)$ with respect to $\{\sigma_j\}$. 
Notably, this kind of maximum-likelihood approach usually requires large amounts of data. 
Consequently, one should either use a single, global value of $\sigma$ for all environments in the crystal, or use the benchmark values to define a prior distribution for $\sigma_j$.
In the following we discuss results obtained using a single, global value of $\sigma$ per chemical species.
The uncertainty in the predicted shifts arising from the conversion of the chemical shieldings by shifting the mean of the GIPAW data to match experiments is generally insignificant and will henceforth be neglected. 

\subsection{\label{subsec:none}Accounting for missing structures}

NMR crystallography currently relies strongly on CSP to generate candidate structures. Although CSP is constantly improving in thoroughness and energetic accuracy, one cannot rule out the possibility that the experimental structure is not among the proposed candidates.
We account for this scenario by adding a virtual structure $\tilde{M}$ to the ensemble of CSP candidates, which represents the ``neglected'' structures. 
While its properties are largely an arbitrary choice, it makes sense to use a Gaussian with a mean and width corresponding to the mean and standard deviation of the shifts of the CSP candidates.
If $\tilde{M}$ has a substantial probability of matching experiment, one should question the comprehensiveness of the CSP candidate pool. 

\subsection{\label{subsec:pca}Visualizing the NMR structural landscape}

Particularly in cases in which the Bayesian analysis does not allow the conclusive identification of the experimental structure, it is useful to gather further insights into the reasons why NMR crystal structure determination has reached the limits of its resolving power, and into whether and how it might be possible to reach a clearer assignment. 
A principal component analysis (PCA) of the shifts of all models provides a means of generating a low-dimensional representation that reflects the similarity of the different models in terms of their NMR shifts, in which one can then embed experiment. 
Unfortunately, for this prior assignments of shifts are required and one is limited to considering shifts from one chemical species. 

We thus instead introduce a universally-applicable approach, based on the definition of a kernel $k(M,M')$, which can be found in appendix~\ref{app:pca} and which reflects the probability that two models could be confused with each other when seen through the lens of their chemical shifts and the available degree of shift-structure assignment.
A kernel PCA (KPCA) extracts a principal component projection of the models (and experiment). This approach owes its universal applicability to the availability of meaningful estimates of $p({\bf y}|M)$ in the presence of shifts from multiple chemical species and irrespective of whether shift assignments are available or not.
Note that, if assignments are indeed available, i.e. when $p({\bf y}|M)$ is defined by Eqs.~(2) and (3), and a global uncertainty $\sigma$ is used, the distances in the KPCA representation again become a direct measure of the shift RMSDs -- with the caveat that distortions can be introduced by the low-dimensional projection.

Embedding the experimentally measured shifts in a low-dimensional representation of the shift similarity provides a scale to the (dis-)similarity of CSP candidates. In cases in which the experimental structure cannot uniquely be identified, it further provides a means of assessing whether two or more models are viable representatives of the experimental structure because they are indistinguishable in terms of their shifts, or because their predicted shifts are too inaccurate to resolve which one agrees with experiment despite distinct shift signatures.

We further perform a PCA on the structural features of all models as described within the smooth overlap of atomic positions (SOAP) framework~\cite{bartok_2013_soap,de_2016_soap}. Loosely speaking, atomic configurations are represented in terms of a smooth atom density, with different chemical species ``tagged'' to be distinguishable from each other~\cite{will+19jcp}. It is constructed as the sum of Gaussian distributions centered on the atomic positions and symmetrised with respect to global translations and rigid rotations of the atomic configuration.
The SOAP features correspond to coefficients obtained by expanding this atom-density description of atomic configurations in spherical harmonics and a set of orthogonal radial basis functions. A more detailed description can be found in section~\ref{sec:mlmodel} and appendix~\ref{app:mlmodel}.
This structural PCA allows us to generate a low-dimensional representation of the structural similarity of the different models. This provides complementary information to the KPCA representation of shift similarity, and permits distinguishing whether a NMR crystal structure determination has reached the limits of its resolving power (a) because structurally dissimilar models produce similar shifts, (b) because the distinction between structurally very similar models is impossible~\cite{hofstetter_2017}, or (c) because the distinction between structurally dissimilar models with dissimilar shifts cannot be made due to the uncertainties in the predicted (and measured) shifts.
It is worth noting that constructing the measure of structural similarity on a SOAP representation of the models is but one particular choice. In general any metric of structural (dis-)similarity -- for example the RMSD-1mol~\cite{chisholm_2005} as used to construct Supplementary Fig.~S17 -- can be used as a basis for a KPCA projection of structural similarity following the approach described in appendix~\ref{app:pca}.

\section{\label{sec:mlmodel}Computational methods}

In section~\ref{sec:applications} we discuss chemical shifts predicted using a ML model, which extends the Gaussian process regression (GPR) model built around the smooth overlap of atomic positions (SOAP) framework~\cite{bartok_2013_soap,de_2016_soap} presented in Ref.~\cite{paruzzo_2018_shiftml} by 
(i) active set sparsification via a projected~process (PP) strategy~\cite{csato_2002_sparsegpr,seeger_2003_sparsegpr,rasmussen_2006_gpr}, 
(ii) the efficient estimation of the uncertainty in predictions using a resampling approach~\cite{musil_2019_uncertainty}, and 
(iii) the radial scaling approach introduced in Ref.~\cite{willatt_2018_rssoap}, which drastically improves the computational performance compared to the original multi-scale approach.
Sparsification of the SOAP descriptions of atomic environments further speeds up predictions.
The construction of the ML model is described in detail in the appendix~\ref{app:mlmodel}.
The new model extends the original ShiftML scheme presented in Ref.~\cite{paruzzo_2018_shiftml} by incorporating sulfur-containing compounds thereby increasing the training set from 2000 to 2500 structures, and (slightly) outperforming it. Crucially, the expected errors of 0.48\,ppm for out-of-sample predictions of $^1$H shifts are comparable to the inherent error of the underlying GIPAW predictions with respect to experiment of around $0.33\pm0.16$\,ppm~\cite{salager_2010_nmr}.

\begin{figure}[tbhp][p]
    \centering
    \includegraphics[width=\linewidth]{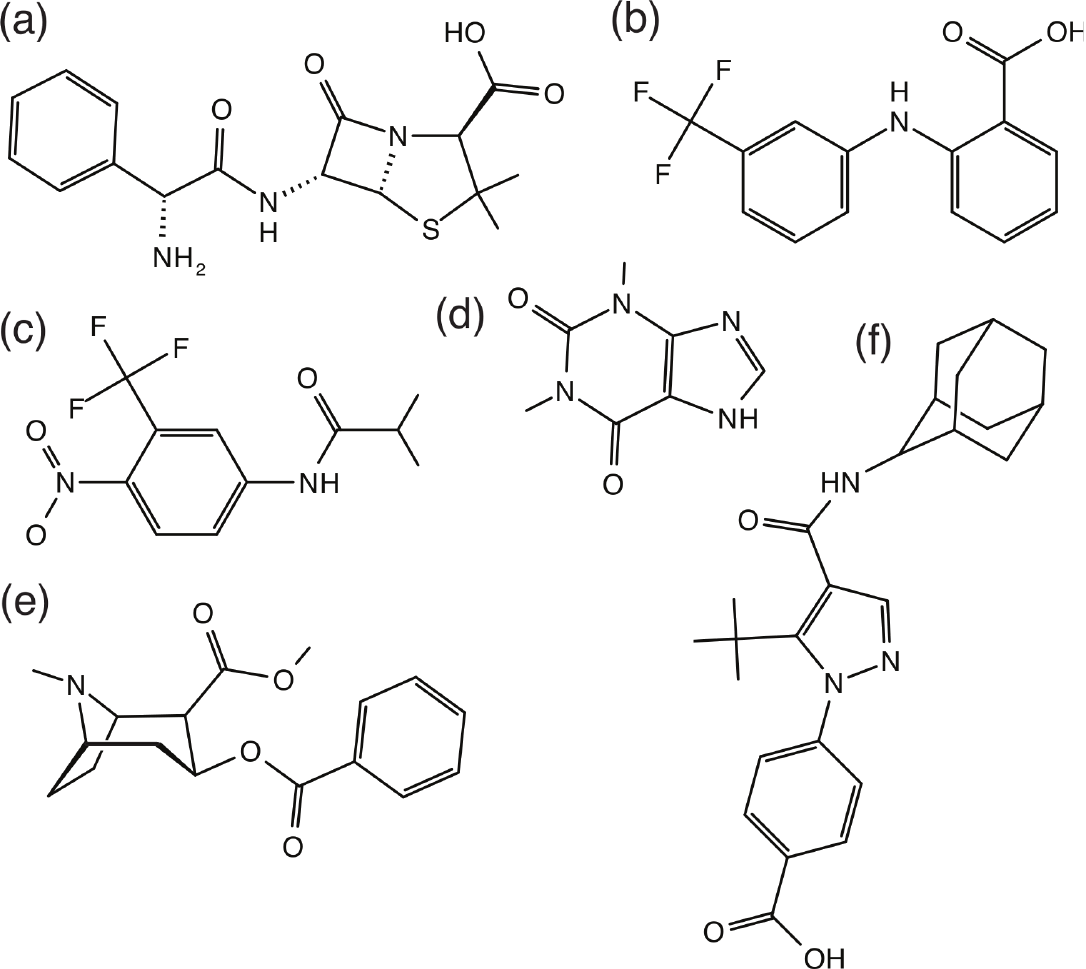}
    \caption{Chemical structures of (a) ampicillin, (b) flutamide, (c) cocaine, (d) theophylline, (e) AZD8329 and (f) flufenamic acid.}
    \label{fig:struct}
\end{figure}
\begin{figure}[tbhp][p]
    \centering
    (a)\phantom{aaaaaaaaaaaaaaaaaaaaaaaaaaaaaaaaaaaaaaaa}\\
    \includegraphics[width=0.675\linewidth]{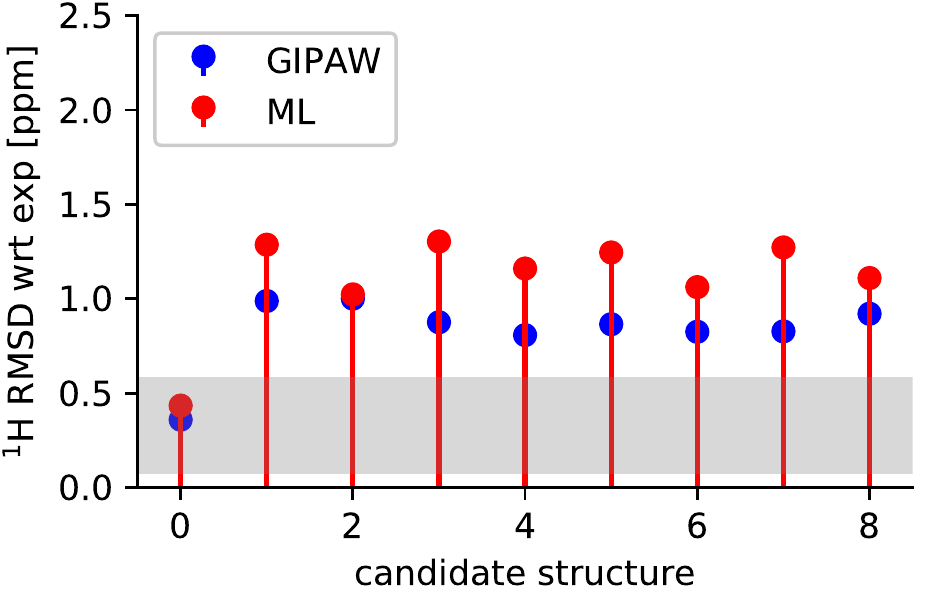}
    (b)\phantom{aaaaaaaaaaaaaaaaaaaaaaaaaaaaaaaaaaaaaaaa}\\
    \includegraphics[width=0.675\linewidth]{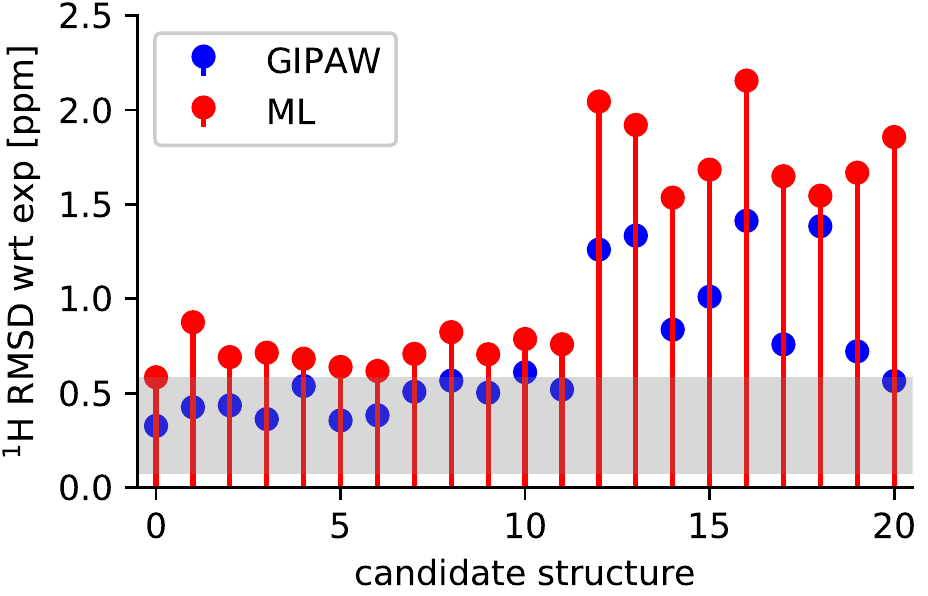}
    \caption{(a) and (b) show the RMSDs of the GIPAW (blue) and ML (red) $^1$H shifts of the AZD8329 and ampicillin CSP candidates with respect to experiment. The gray area indicates the one sigma confidence interval for the GIPAW $^1$H shifts as determined by the typical error of GIPAW predictions with respect to the experimentally measured shifts for a set of benchmark compounds of known atomic structure~\cite{salager_2010_nmr,hartman_2016,dracinsky_2019}.}
    \label{fig:rmsds}
\end{figure}

It is worth noting that Liu \textit{et al.} have recently demonstrated that, despite replacing the SOAP description of atomic densities with a non-symmetry-adapted real-space discretised equivalent, a sufficiently complex neural network architecture can tease out improvements of up to around 20\% in prediction accuracy using the original training data~\cite{liu_2019_densenet}. 
We nonetheless here choose a SOAP-GPR framework noting that 
the statistical ML uncertainties are uncorrelated with the inherent errors of the reference GIPAW data and must therefore be added to the GIPAW error(s) in quadrature.
In consequence, reductions in ML errors at this point reap insignificant improvements to the resolving power of ML-based NMR crystallography without accompanying reductions in the underlying GIPAW errors with respect to experiment.
The SOAP-GPR framework is robust, easily trained, has recently been generalised to the prediction of tensorial properties such as (anisotropic) chemical shielding tensors~\cite{grisafi_2018_tensorial}. Furthermore, it provides accurate estimates of prediction uncertainty~\cite{musil_2019_uncertainty}. These are particularly important in this context, not only to estimate the reliability of assignments, but also because GIPAW calculations can at times yield unreliable results, and the ML model can be improved by automatically discarding problematic training data (see appendix~\ref{app:mlmodel}).

\begin{figure*}[bth]
    \centering
    \begin{minipage}[t]{0.68\linewidth}\vspace{0pt}
    \includegraphics[width=\textwidth]{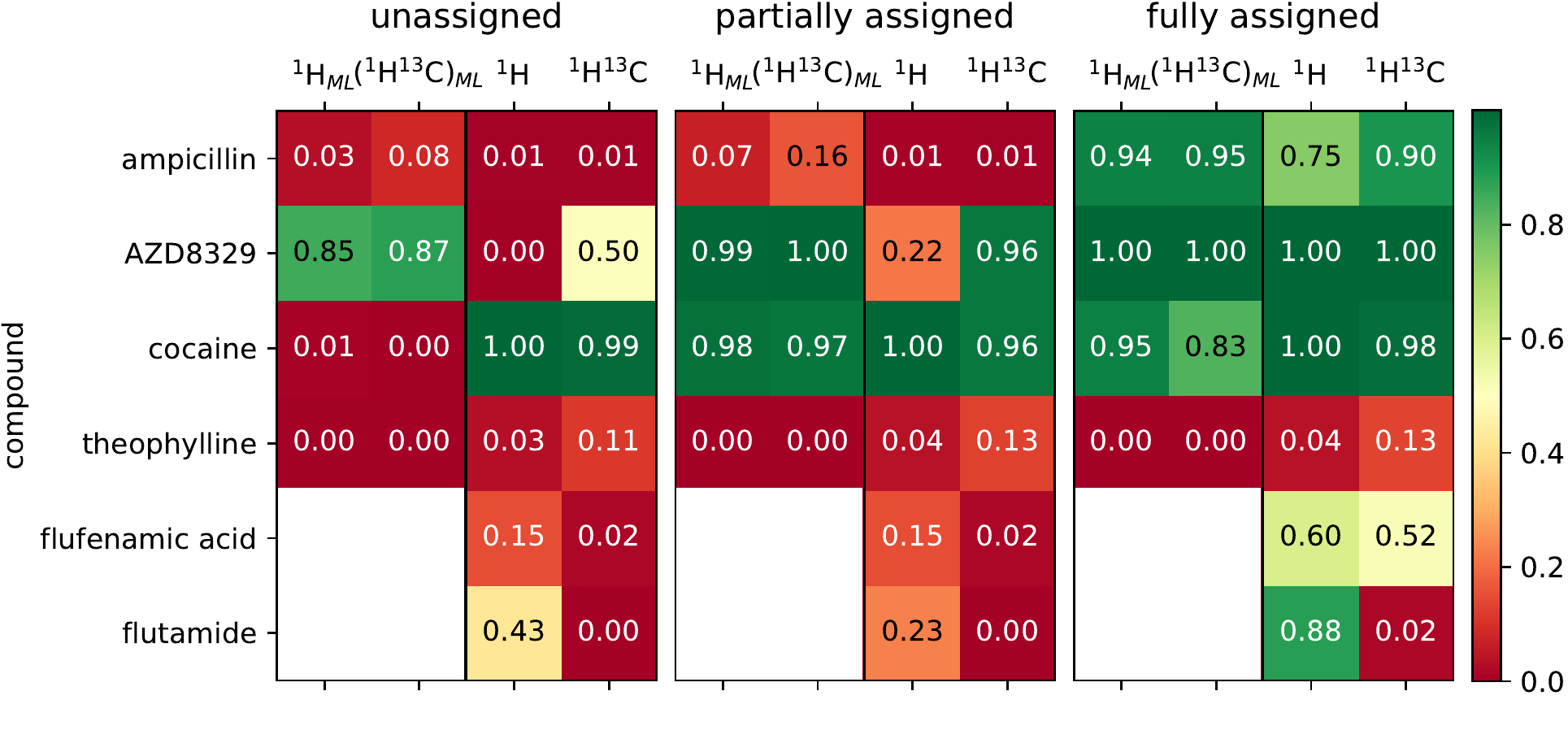}
    \end{minipage}
    \hfill
    \begin{minipage}[t]{0.30\linewidth}\vspace{0pt}
    \caption{Overview of the results of NMR crystal structure determinations for the six benchmark compounds based on different degrees of assignments of the experimental shifts to nuclei and using $^1$H and $^{13}$C shifts calculated with ML or GIPAW, respectively. Both full (fully assigned) and partial assignments (partially assigned) are detailed in Supplementary Section~SII.
    Each cell is colored and labeled according to the Bayesian probability of matching experiment assigned to the representative of the experimental structure among the CSP candidates -- this probability provides the key indicator of the reliability of the structure determination.}
    \label{fig:overview}
    \end{minipage}
\end{figure*}
\section{\label{sec:applications}Results and discussion}

In order demonstrate the Bayesian approach to NMR crystallography, we use it to quantify the confidence in the structure determination of six molecular crystals (see Fig.~\ref{fig:struct}). 
We also demonstrate the use of two-dimensional visualisations of the similarity between candidate structures, both in terms of their structural features and in terms of their predicted chemical shifts, following the recipe of section~\ref{subsec:pca}.

\subsection{Benchmark systems}

Ampicillin, 4-[4-(2-adamantylcarbamoyl)-5-tert-butyl-pyrazol-1-Yl] benzoic acid (referred to as AZD8329), cocaine, theophylline, flufenamic acid, and flutamide (depicted in Fig.~\ref{fig:struct}) have all previously been studied using NMR crystallography~\cite{hofstetter_2019_nmr,baias_2013_nmr,baias_2013_csp,pinon_2015}.
In each case the experimental NMR shifts have been fully assigned to nuclei, the corresponding crystal structures are known, and GIPAW shifts for a pool of CSP candidates are available. 
Furthermore, for all six compounds the CSP candidates include a representative of the experimental structure, which is referred to as the correct candidate in the following.
(The partial and full assignments of the experimentally measured shifts to particular nuclei in the compounds used in the following are detailed in Supplementary Section~SII.)

Fig.~\ref{fig:rmsds} shows examples of the analysis that is traditionally performed in NMR crystallography. The RMSD between the experimental shifts and those predicted for multiple CSP candidates is computed using fully assigned $^1$H shifts, and compared to the typical uncertainty of GIPAW (or ML) predictions. 
The structure with the lowest RSMD is deemed to be the best candidate and identified as the experimental structure, provided the RMSD is consistent with the inherent uncertainty in the predicted shifts. 
In the case of AZD8329, only one structure is consistent with experiment, making the structure determination conclusive. 
In the case of ampicillin, although the correct candidate has the lowest RMSD, several others are consistent with experiment within the inherent uncertainty in their predicted shifts. Based on this analysis, it is consequently impossible to assess how trustworthy identifying the best candidate as the experimental structure would be.
In practice energetic considerations strongly favour the correct candidate and facilitate determining the correct crystal structure.

\begin{figure*}
    \centering
    \begin{tabular}{lll}
        (a) & (b) & (c)\\
        \includegraphics[width=0.315\textwidth]{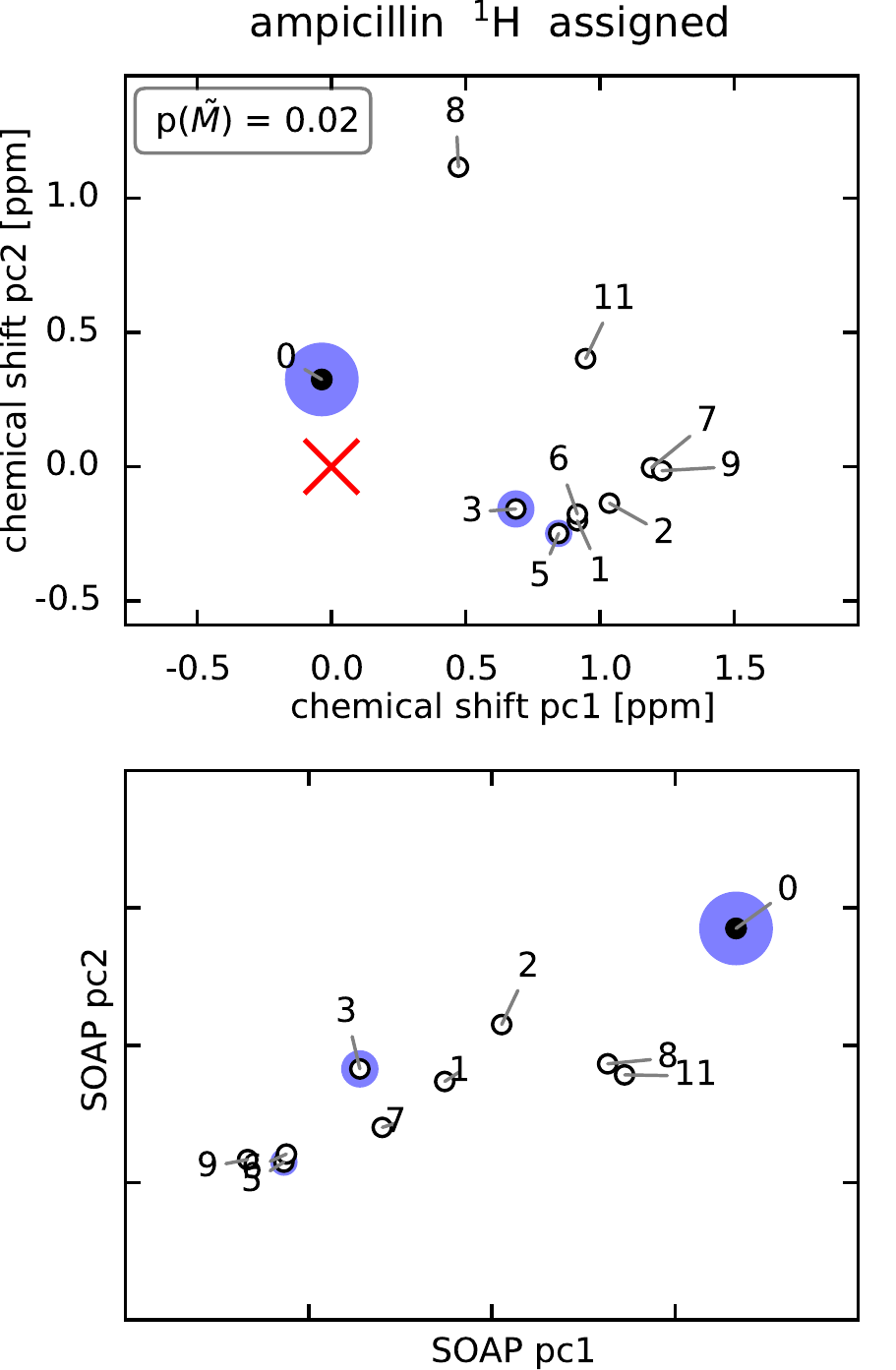} & 
        \includegraphics[width=0.315\textwidth]{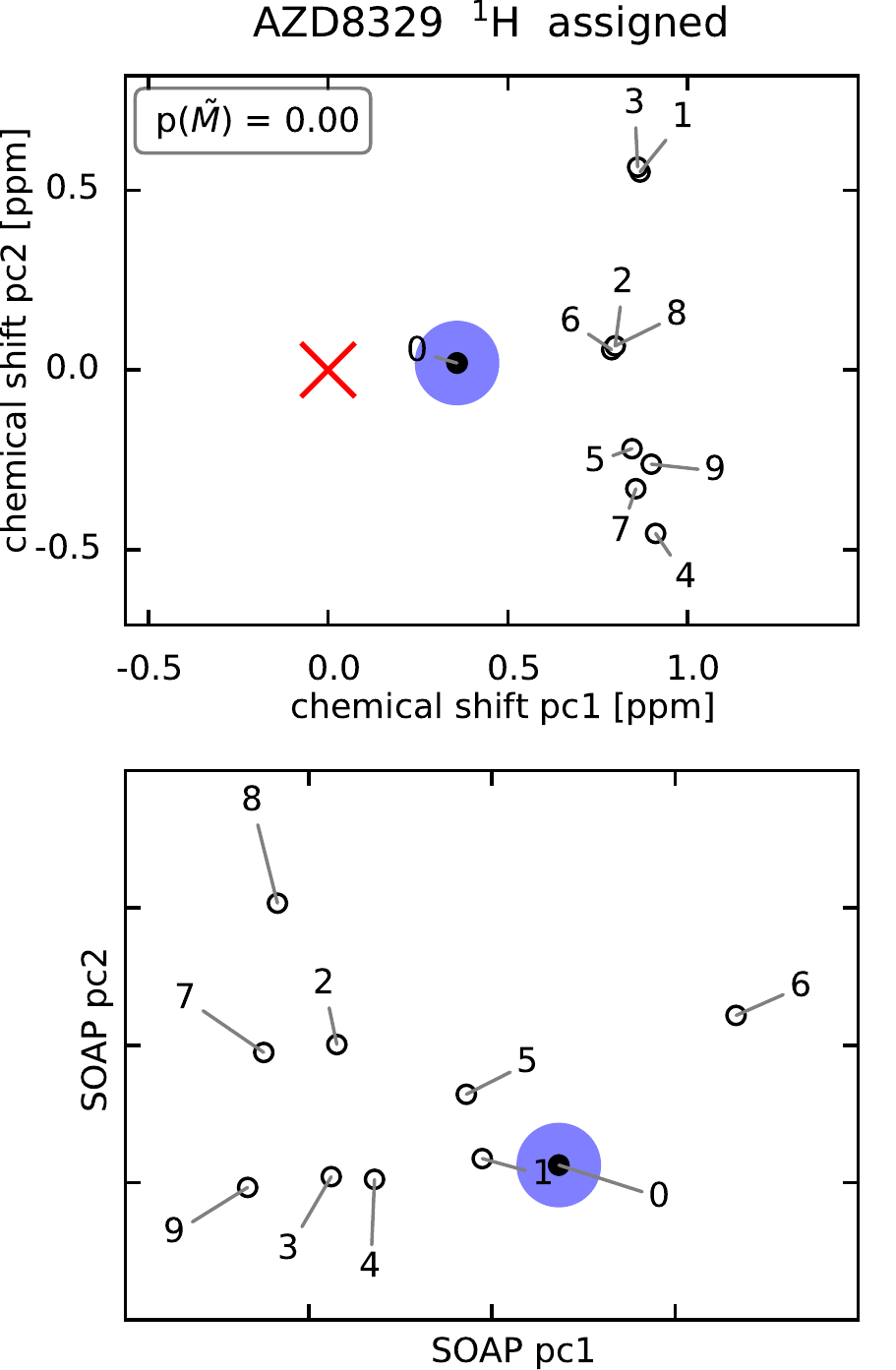} & 
        \includegraphics[width=0.315\textwidth]{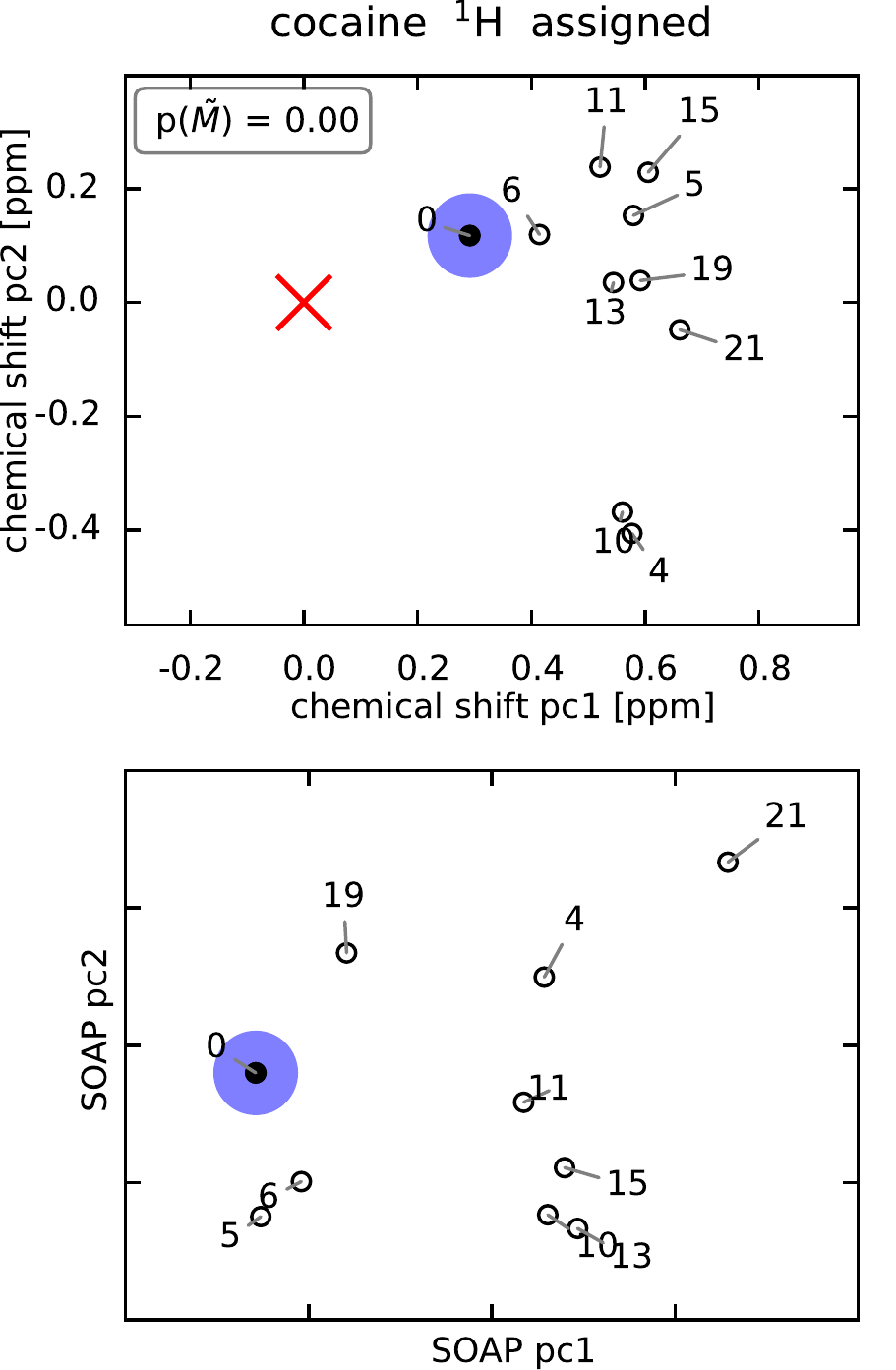} \\
        (d) & (e) & (f)\\
        \includegraphics[width=0.315\textwidth]{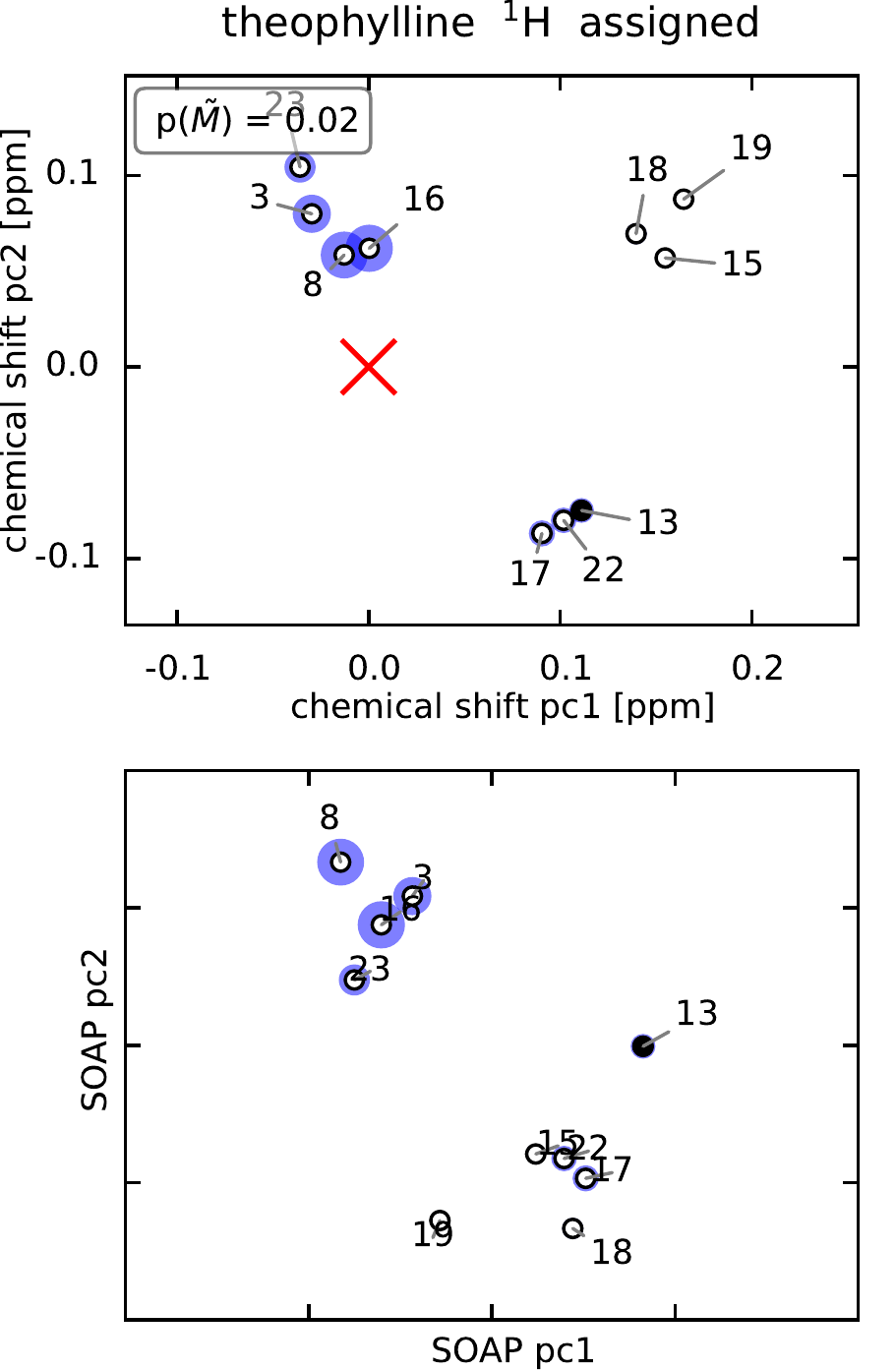} & 
        \includegraphics[width=0.315\textwidth]{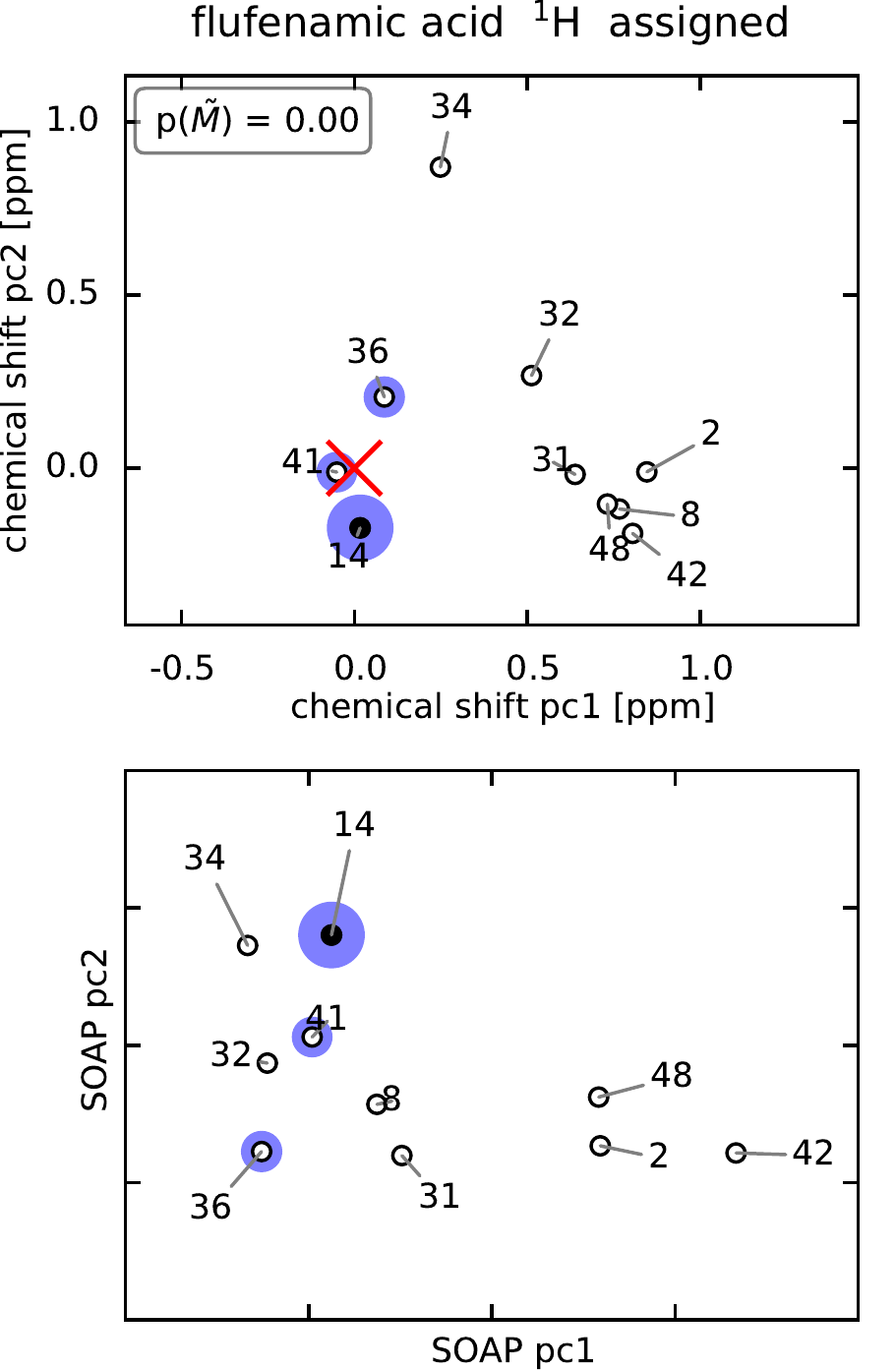} & 
        \includegraphics[width=0.315\textwidth]{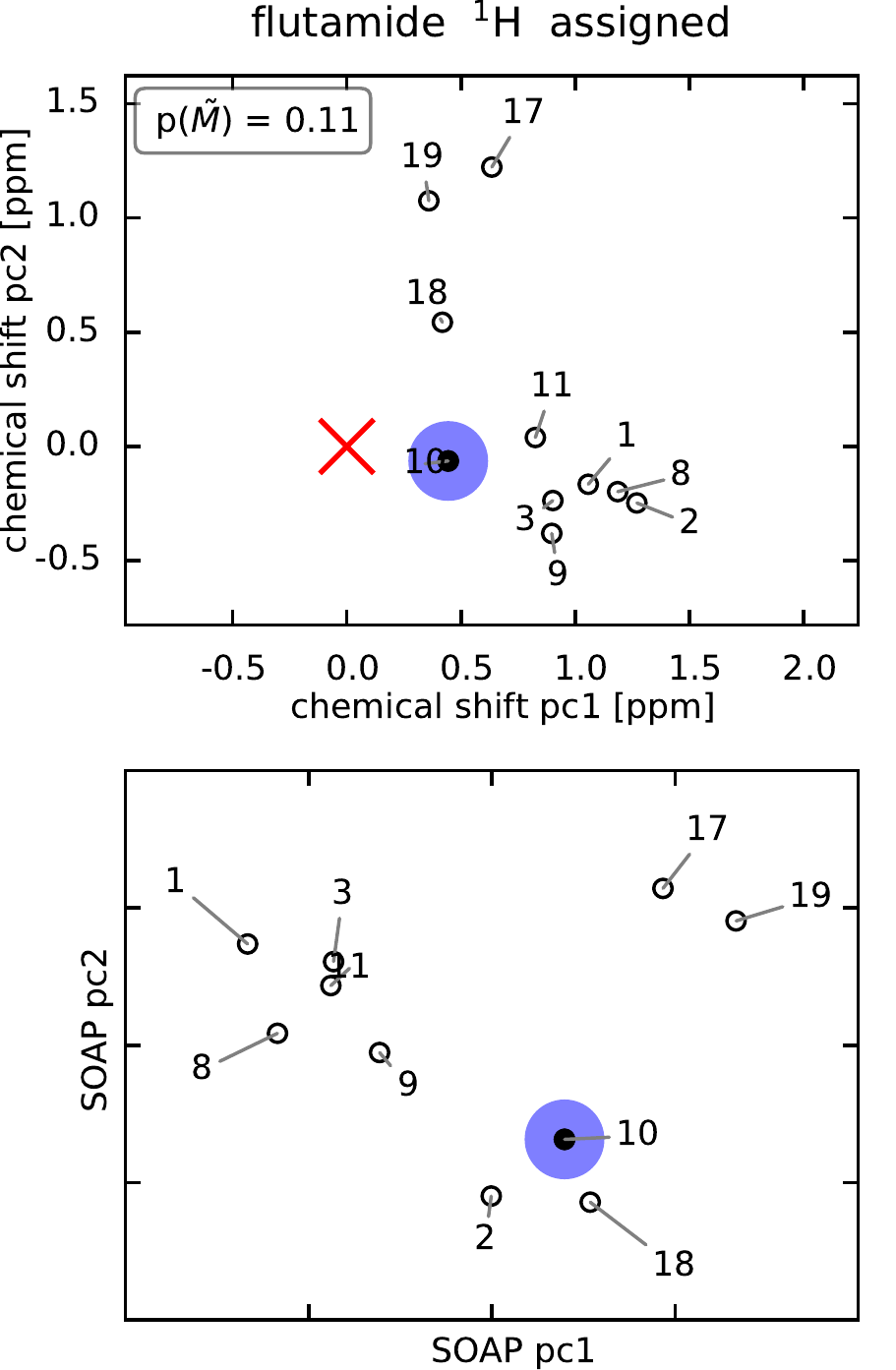} \\
    \end{tabular}
    \caption{Evaluation of the top 10 (a) ampicillin, (b) AZD8329, (c) cocaine, (d) theophylline, (e) flufenamic acid, and (f) flutamide CSP candidates. 
    The correct candidates are shown as filled circles and the others as empty circles. 
    For each candidate the probability of matching experiment $p\pcond{M}{\by^\star}$ is indicated by the area of the blue disk.
    The candidates are labelled according to their rank in terms of configurational energy with zero indicating the energetically most favourable candidate.
    The respective upper panels show the similarity of the candidates to each other and to the (out-of-sample embedded) experimental data (shown as a red cross) in terms of their fully assigned $^1$H GIPAW shifts. $p(\tilde{M})$ denotes the probability that the virtual candidate, which represents structures potentially missing from the CSP candidate pool, matches experiment.
    The respective lower panels show the structural similarity of the candidates in terms of their SOAP features.
    While the relative distances of structures are a measure of their (dis-)similarity, the absolute value of the principal components (pc) from the (K)PCA constructions described in section~\ref{subsec:pca} has no intuitive physical meaning and is therefore not shown.}
    \label{fig:similarity_overview}
\end{figure*}
\subsection{\label{subsec:results_full}Quantitative structure determination and visualisation}

Cases such as ampicillin, in which NMR structure determinations are complicated by the presence of two or more candidates in close agreement with experimental NMR shift data, are the primary reason for developing the Bayesian framework discussed in section~\ref{sec:theory}. 
From Fig.~\ref{fig:overview} we see that on the basis of the same $^1$H shifts from GIPAW calculations, we estimate that the correct structure is identified with confidence in 4 out of the 6 benchmark cases (75\% for ampicillin, 88\% for flutamide, and 100\% for AZD8329 and cocaine), and with some uncertainty in the case of flufenamic acid (60\%). In the case of theophylline, the analysis confirms that the experimental structure cannot be distinguished (see Fig.~\ref{fig:overview} and Ref.~\cite{baias_2013_csp}). 

In order to elucidate why the level of confidence in the structural determination varies among the benchmark problems, we generate a two dimensional visualisation in which the CSP candidates for each compound are arranged such that pairwise distances reflect their dissimilarity, and which simultaneously shows the probability with which each candidate matches experiment.
Fig.~\ref{fig:similarity_overview} shows the representations of the similarity of the CSP candidates for each of the six compounds. 
For each compound we show the similarity in terms of $^1$H chemical shifts (top panels) and in terms of structure (lower panel).

The similarity in terms of chemical shifts reflects the resolving power of NMR. The similarity in terms of their structural features reflects how distinct the geometries of different candidates are.
By embedding experiment, i.e. the experimentally measured shifts, in the representations of shift similarity one can also assess how closely (or not) the shifts of different candidates agree with experiment. 
First, by looking at the similarity as seen through the chemical shifts one can tell whether failure to identify conclusively the correct structure is due to lack of resolving power of NMR, or to the inaccuracy of the predicted shifts. 
For example, the case of theophilline (Fig.~\ref{fig:similarity_overview} (d)) shows that structures 8 and 16, which are identified as the most likely candidates, exhibit very distinct $^1$H shifts from structure 13, which is the correct candidate. Hence, even though there are only four $^1$H shifts, this analysis suggests that more accurate predictions of the $^1$H shifts would probably suffice to correctly determine the structure. 
In contrast, in the case of flufenamic acid (Fig.~\ref{fig:similarity_overview} (e)) the three structures with non-zero probability are all similarly close to experiment as they are to each other\footnote{Actually, the Fig.~\ref{fig:similarity_overview} (e) seems to indicate that structure 41 is closer to experiment than structure 14, whose chemical shifts agree most closely with the experimentally measured ones and which happens to be the correct candidate. This distortion is an artifact of the projection of the NMR (and geometric) similarities, which correspond to a distance in a high-dimensional space, onto a two-dimensional representation.}. In this case, it seems that shifts from additional chemical species, or a dramatic increase in the accuracy of shift predictions, would be needed to resolve the ambiguity. 

Whenever two or more structures are close together in the shift-based representation, it would be hard to distinguish them by means of a NMR experiment. For instance, this is the case for structures 13, 17 and 22 of theophylline, as can be seen in Fig.~\ref{fig:similarity_overview} (d). 
Meanwhile, the geometry-based representation, which is also shown in Fig.~\ref{fig:similarity_overview} (d), clearly shows that structure 13 is actually distinct. This geometric difference is not reflected in the value of the shifts, which is at least in part due to the small number of hydrogen atoms in a theophylline molecule. 
For comparison, the similarity of the structures 3, 8, 16, 23 in terms of chemical shifts clearly reflects an underlying geometric similarity.

\subsection{The importance of experimental assignments of shifts}

While the scenario of NMR crystal structure determinations on the basis of fully assigned $^1$H shifts is the default in modern day NMR crystallography, it is important to appreciate that in general full assignments of chemical shifts to specific nuclei may not be available.
In order to assess the importance of experimental assignments of shifts, we have further evaluated the confidence of structure determinations for each of the six compounds in the partial or complete absence of shift assignments.
These data are shown in Fig.~\ref{fig:overview}, which visualises the confidence with which the correct candidate can be identified as matching the experimental structure (as the key indicator of the \textit{a posteriori} reliability of each structure determination) for each compound and each scenario.
Unsurprisingly, the figure emphasises that experimental assignments of shifts are invaluable.
\begin{figure}[tbhp]
    \centering
    \includegraphics[width=0.34\textwidth]{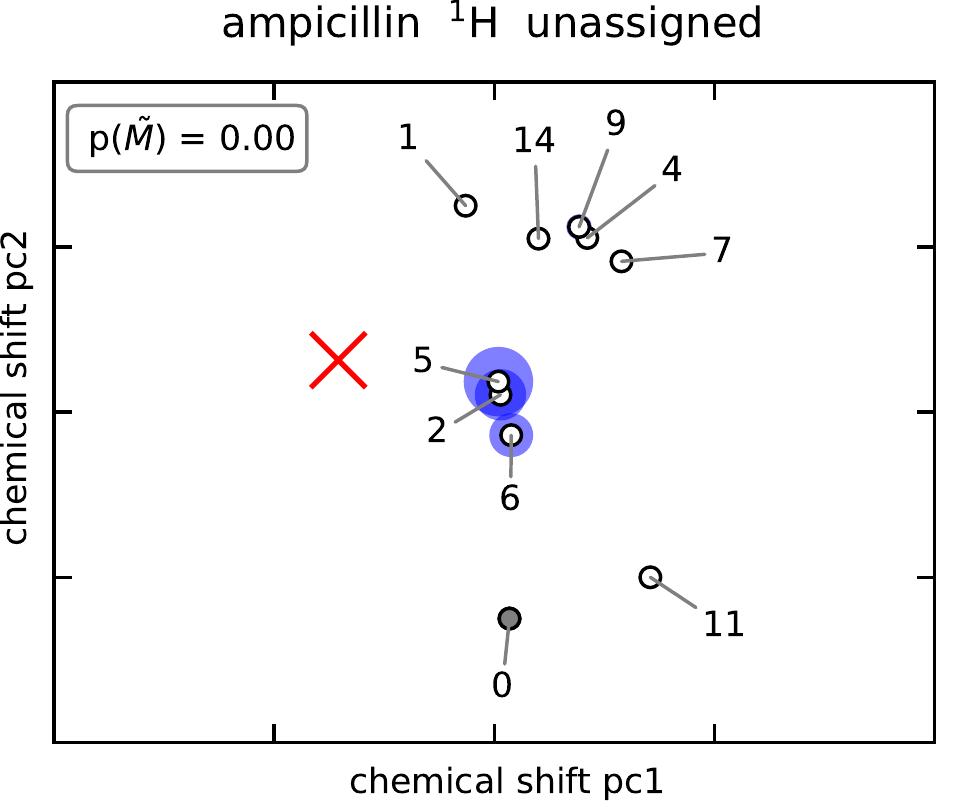}
    \caption{Similarity of the top 10 ampicillin candidates in terms of their unassigned $^1$H GIPAW shifts. While the correct candidate is not among the probable candidates in the absence of shift assignments, it has a fingerprint of $^1$H shifts, which is unlike any other candidate. %
    }
    \label{fig:similarity_ampicillin}
\end{figure}
For instance, the structure determination for ampicillin, which is all but impossible in the absence of experimental assignments of shifts, can be made with 75\% confidence on the basis of fully assigned $^1$H shifts.
Interestingly, for this case Fig.~\ref{fig:similarity_ampicillin} shows that the incorrect structure determination in the absence of experimental assignments is not due to intrinsic lack of resolving power. The correct structure stands out clearly from the others, and as a consequence more accurate $^1$H shifts would suffice to unambiguously identify it as the experimental structure, motivating efforts towards more accurate predictions of NMR shifts beyond GIPAW calculations.

\begin{figure}[tbhp]
    \centering
    (a)\phantom{aaaaaaaaaaaaaaaaaaaaaaaaaaaaaaa}\\
    \includegraphics[width=0.34\textwidth]{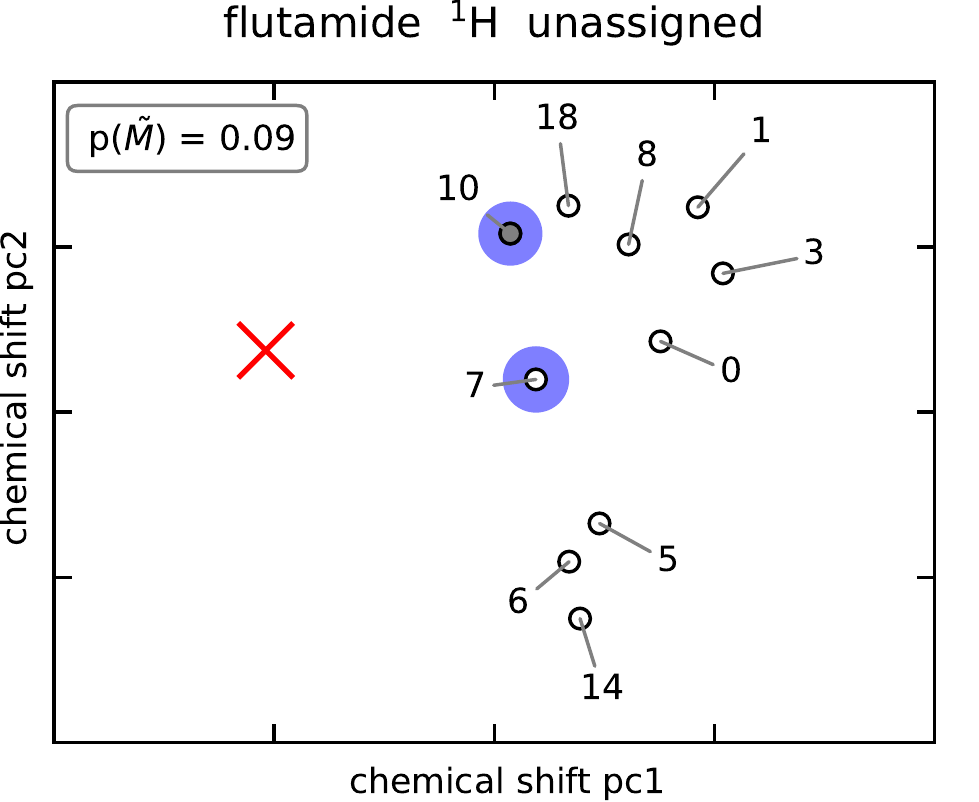}\\
    (b)\phantom{aaaaaaaaaaaaaaaaaaaaaaaaaaaaaaa}\\
    \includegraphics[width=0.34\textwidth]{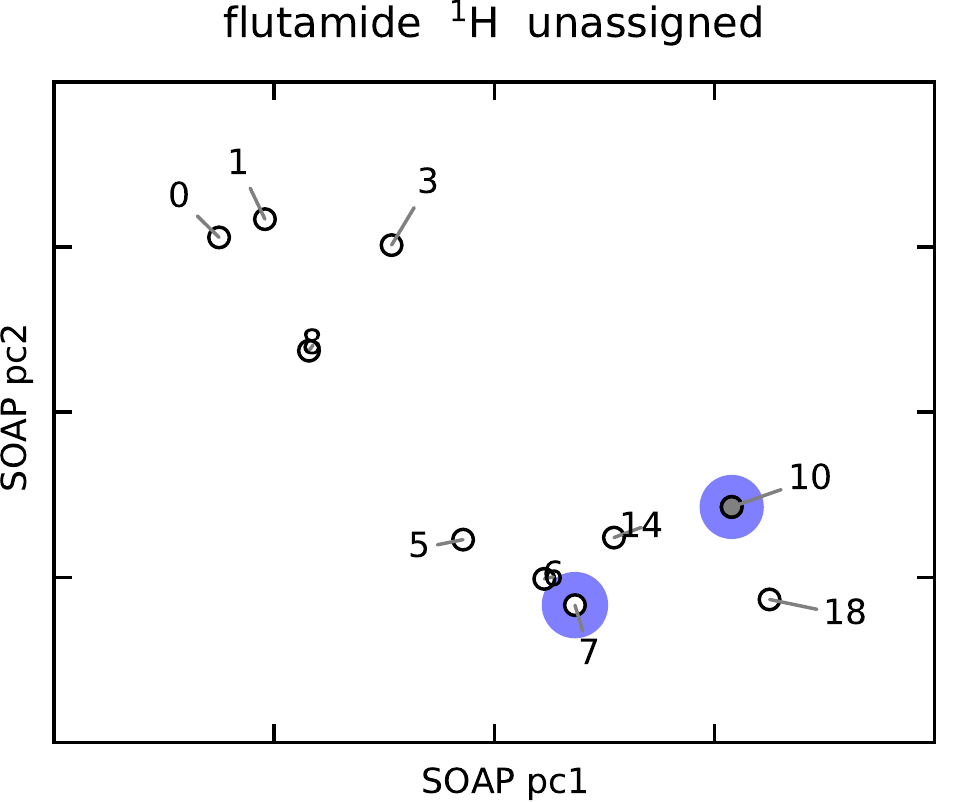}
    \caption{Similarity of the top 10 flutamide candidates in terms of their unassigned $^1$H GIPAW shifts. Candidates 7 and 10 (the correct candidate) both agree similarly well with experiment so that experimental assignment of the shifts are required to determine the experimental structure (see Fig.~\ref{fig:similarity_overview}). Notably candidates 7 and 10 are sufficiently distinct -- both structurally (see panel (b)) and in terms of the $^1$H shifts -- that more accurate predictions of $^1$H shifts would likely allow the determination of the experimental structure even in the absence of shift assignments.}
    \label{fig:similarity_flutamide}
\end{figure}
Shift assignments are similarly critical to achieve structure determination for flutamide with 88\% confidence.
In their absence a second flutamide structure exhibits a very similar NMR signature to the correct candidate (see Fig.~\ref{fig:similarity_flutamide} (a)) and is consequently in sufficiently good agreement with the experimental data to prevent identification of the experimental structure. Notably, the representation of structural similarity shown in Fig.~\ref{fig:similarity_flutamide} (b) indicates that this competing candidate is also structurally similar to the correct candidate.
Yet, in this case experimental shift assignments suffice to resolves the subtle differences between the two competing candidates (see Fig.~\ref{fig:similarity_overview} (b)).
The case of flutamide also evidences the value of introducing the virtual structure $\tilde{M}$ as a representative of configurations missing from the candidate pool. For flutamide it acquires a significant probability of best representing experiment, suggesting noticeable structural differences between the correct candidate and experiment. Yet, the differences between the correct candidate and the XRD structure are very subtle (the atomic positions agree with the structure determined by single crystal X-ray diffraction to within less than $0.1 \ang$), hinting at the role of nuclear motion, which is not considered here (and generally only rarely considered in simulations~\cite{dumez_2009,robinson_2010,dracinsky_2013,dracinsky_2014,monserrat_2014,hass+12jacs}).

\subsection{NMR crystallography using ML predictions of chemical shifts}

Above we have made use of extensive preexisting GIPAW NMR calculations. 
In practice GIPAW shift predictions come at substantial cost, if the size and complexity of the system of interest permits them in the first place.
Fortunately ML shift predictions prove sufficiently reliable to determine structures. 
This is demonstrated by reconstructing the Bayesian models on ML shifts for all systems except flufenamic acid and flutamide. The latter two contain fluorine, leaving them outside the scope of the current ShiftML model.
The results are shown in Fig.~\ref{fig:overview} and demonstrate that ML-based NMR crystallography almost matches the resolving power achieved with GIPAW predictions of NMR shifts.

\begin{figure}[tbhp]
    \centering
    \begin{tabular}{ll}
    (a) & \\
        & \includegraphics[width=0.34\textwidth]{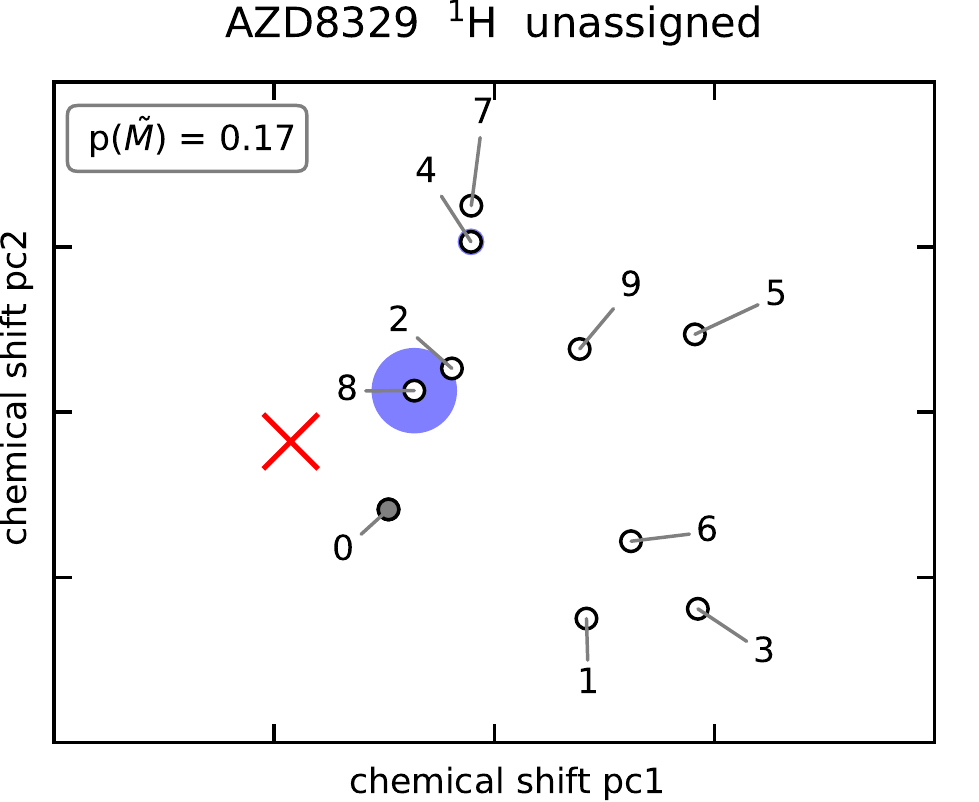}\\
    (b) & \\
        &\includegraphics[width=0.34\textwidth]{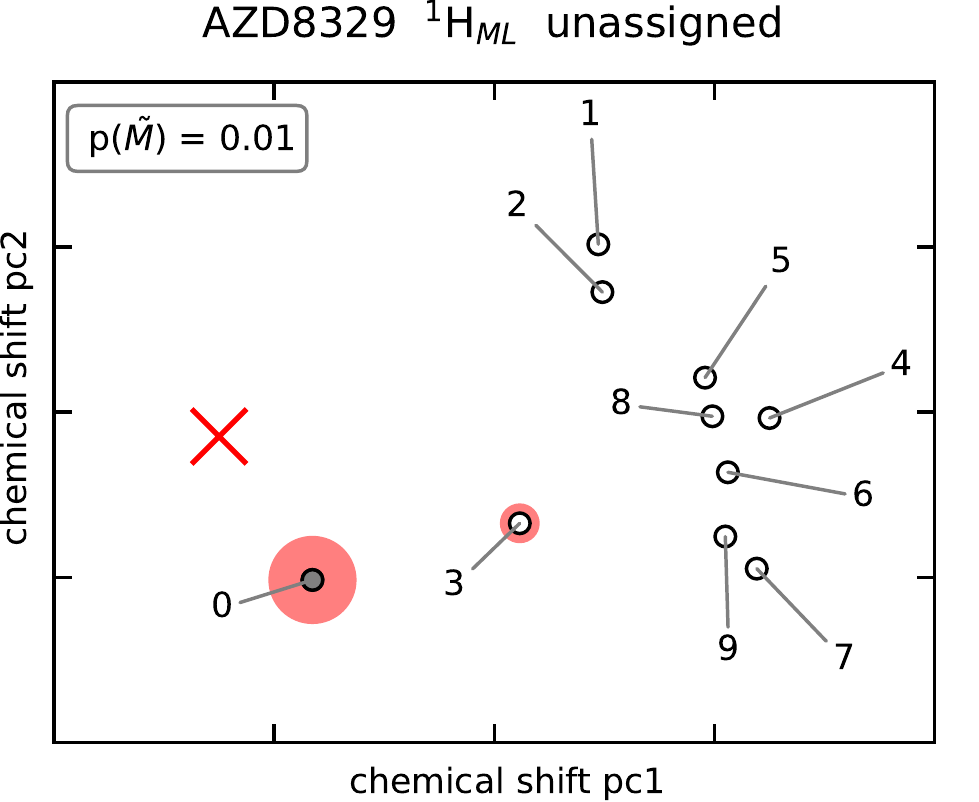}
    \end{tabular}
    \caption{Similarity of the top 10 AZD8329 candidates in terms of their unassigned $^1$H (a) GIPAW and (b) ML shifts. Panel (b) shows that, in terms of the ML shifts, the correct candidate is very distinct from all other candidates and uniquely similar to experiment, even in the absence of experimental shift assignments. In contrast, in terms of the GIPAW shifts the correct candidate is much less distinguishable from the other candidates and is not identified as a probable match with experiment.}
    \label{fig:similarity_azd}
\end{figure}

The case of AZD8329 deserves further discussion. Barring fortuitous cancellation of errors as in the case of ampicillin with fully assigned shifts, the resolving power of ML-based NMR crystallography can only approximate but not surpass that achieved on the basis of shifts calculated using the reference method. 
As we discuss in Appendix~\ref{app:mlmodel}, however, GIPAW calculations are subject to instabilities, which are reflected in the occasional appearance of discrepancies between ML and GIPAW predictions that are more than three times larger than the corresponding uncertainty in the ML prediction. 
The fact that he accuracy of the ML model improves by dropping these structures from the training underscores the fact that this discrepoancy reflects inconsistencies in the reference calculations.

For AZD8329 the GIPAW shifts for the correct candidate are consistent with the ML predictions to within the estimated ML uncertainties (explaining the small observed shift RMSD in Fig.~\ref{fig:rmsds} (a)), but the GIPAW shifts of many of the ``false'' candidates would be classified as outliers according to the scheme we use to eliminate unstable GIPAW calculations from the training set.
Consequently, the structure determination based on GIPAW shifts requires full assignments, whereas the ML shifts (which are not tainted by instabilities) allow for the correct determination of the structure even when assignments are incomplete or entirely unavailable (see Figs.~\ref{fig:overview} and \ref{fig:similarity_azd}).
This additional resolving power is also present in the fully assigned case, but since the GIPAW shifts are already sufficient to determine the structure with 100\% confidence, the improvement with ML is not visible.

\subsection{$^{13}$C NMR crystallography}

Irrespective of whether NMR shifts are predicted using GIPAW calculations or ML methods, $^1$H shifts do not always suffice to pin down the experimental structure.
The cases of flufenamic acid and theophylline highlight the limits of $^1$H NMR crystallography for compounds with few distinct hydrogen atoms, with a low, 60\% confidence in the structure determination in the former case, and the determination of the experimental structure being simply impossible in the latter. 
This makes it tempting to turn to $^{13}$C data in search for more information to exploit in distinguishing the experimental structure.
However, in agreement with current wisdom~\cite{baias_2013_nmr}, Fig.~\ref{fig:overview} suggests that the inclusion of $^{13}$C shifts reduces the confidence in the identification of the experimental structure.
The fact that the resolving power of NMR crystallography appears to deteriorate upon inclusion of $^{13}$C shifts warrants further discussion. {In a Bayesian framework, adding more information should never degrade the prediction accuracy, unless the accuracy of such information is overestimated.} 
The degradation of prediction accuracy therefore indicates that the value of $\sigma_{\ce{C}}^\text{DFT} = 1.9 \pm 0.4$\,ppm based on benchmark data~\cite{salager_2010_nmr,hartman_2016,dracinsky_2019} substantially underestimates the actual error for the compounds considered here. 
Following the strategy of maximizing $p(\by^\star)$ proposed in section~\ref{subsec:gipaw_errors}, the typical error in $^{13}$C shifts can be estimated to a substantially larger $2.7 \pm 0.9$\,ppm. This is substantiated by the RMSD of the GIPAW shifts of the correct candidates with respect to the respective experimentally measured shifts of $2.6 \pm 1.4$\,ppm. For comparison, the corresponding RMSD of the $^1$H GIPAW shifts is $0.28 \pm 0.09$\,ppm and thus entirely consistent with the global estimate of $\sigma_{\ce{H}}^\text{DFT} = 0.33 \pm 0.16$\,ppm.

\begin{figure}[tbhp]
    \centering
    \includegraphics[width=0.35\textwidth]{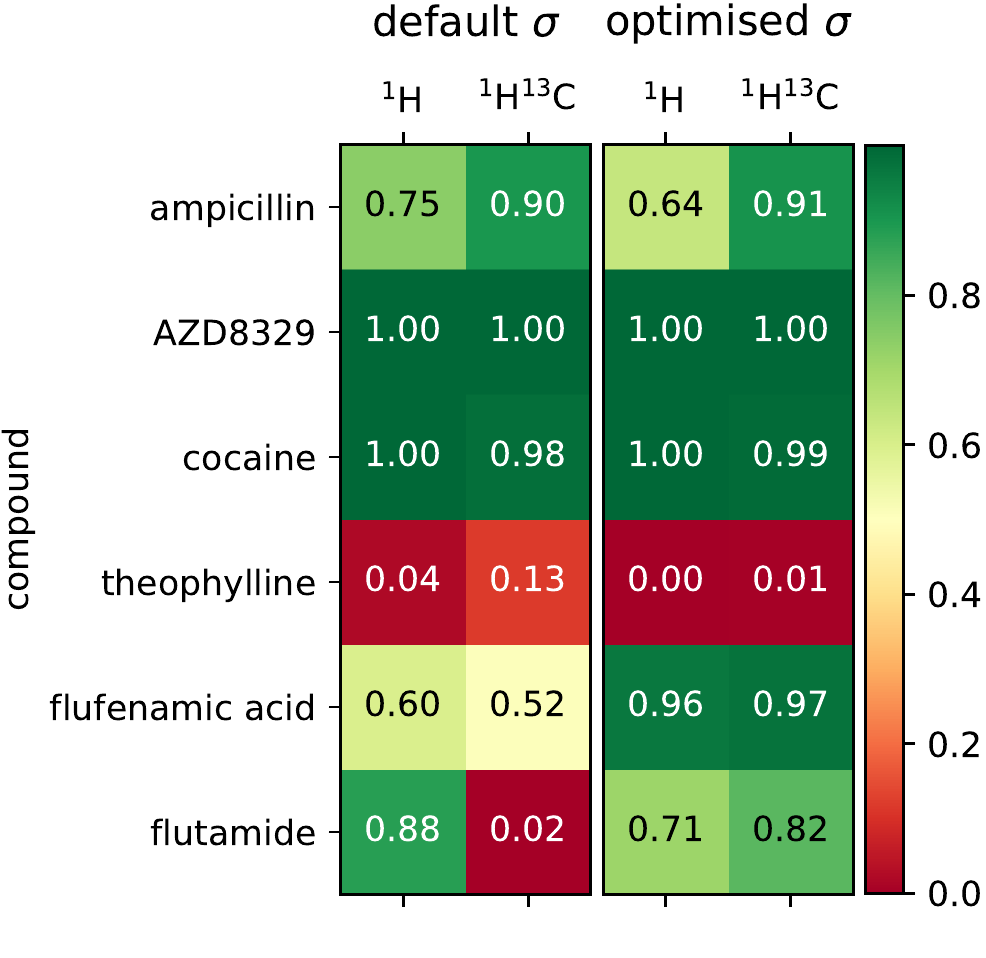}
    \caption{Comparison of the Bayesian probabilities of matching experiment assigned to the representative of the experimental structure among the CSP candidates on the basis of the default global uncertainties $\sigma_{\ce{H}}^\text{DFT} = 0.33 \pm 0.16$\,ppm and $\sigma_{\ce{C}}^\text{DFT} = 1.9 \pm 0.4$\,ppm (left) and uncertainties $\sigma_{\ce{H}} = 0.28 \pm 0.09$\,ppm and $\sigma_{\ce{C}} = 2.7 \pm 0.9$\,ppm estimated for each individual compound under consideration by maximizing $p(\by^\star)$ with respect to $\{\sigma_j\}$ as described in section~\ref{subsec:gipaw_errors} (right).}
    \label{fig:carbon_shift_errors}
\end{figure}

Fig.~\ref{fig:carbon_shift_errors} demonstrates that, provided the compound-dependent, data-driven estimate of the errors in GIPAW $^{13}$C shifts derived here is used, the inclusion of $^{13}$C shifts in the analysis indeed tends to improve rather than impair the resolving power of NMR crystallography.
For instance, for flufenamic acid the structure determination is not limited by the accuracy of the predicted $^1$H (and indeed $^{13}$C) shifts, but rather by the accuracy of the estimates of the typical errors in those shift. \textit{Accordingly, its structure can be determined with almost complete confidence (96\%) provided accurate estimates of the typical errors in $^1$H (and $^{13}$C) shifts (see Fig.~\ref{fig:carbon_shift_errors}).}

\section{\label{sec:conclusions}Conclusions}

We have introduced an analysis framework for NMR crystal structure determination, which is suited to a variety of experimental (and computational) setups. 
By quantifying the confidence in identifications of experimental structures our analysis framework demonstrates that definitive identifications are sometimes possible even if the corresponding shift RMSD does not fall within the traditional ``confidence interval''. This relies on exploiting all available information, much of which the traditional RMSD measure of agreement with experiment is blind to.
Notably, we use this approach to conclude that literature benchmarks for the accuracy in the prediction of $^{13}$C chemical shifts underestimate the uncertainties. 
We find that $^{13}$C errors for GIPAW predicted shifts for the compounds used here are $2.7 \pm 0.9$\,ppm as opposed to previous,  estimates of $1.9 \pm 0.4$\,ppm. If we use our corrected error estimates, incorporating $^{13}$C shifts into the analysis improves the reliability of structure determination. In one of the cases we considered, the use of self-consistently computed uncertainties lifts the ambiguity on the structure determination.

We also introduce a visual representation of the crystal structure landscape based on a low-dimensional projection that reflects the similarity between the structure of the candidates, or directly on their NMR shifts. These visualisations help determine whether lack of structural diversity, insufficient resolving power of the experiment, or uncertainties in the computationally-determined shifts are involved in inconclusive structural determinations. 
In combination, the Bayesian framework and the low-dimensional representations of candidate similarity provide an integrated way of 
(1) identifying  among a pool of candidate structures which most closely approximates the experimental one, 
(2) performing sanity checks of the comprehensiveness of the pool, the associated predicted NMR shifts, and the initial identification, 
(3) quantifying the confidence in the identification assuming the sanity checks have provided satisfactory results, 
(4) analyzing what factors limit the confidence or, when definitive identification of the experimental structure is not possible, the resolving power of the crystal structure determination.
\begin{acknowledgments}
EAE, AA, and MC acknowledge funding by the European Research Council under the European Union's Horizon 2020 research and innovation program (grant no. 677013-HBMAP).
EAE acknowledges support by the NCCR MARVEL, funded by the Swiss National Science Foundation (SNSF).
AH, FMP and LE acknowledge financial support from the Swiss National Science Foundation (grant no. 200020\_178860). 
\end{acknowledgments}

\clearpage
\appendix

\section{\label{app:mlmodel}The ShiftML 1.1 model}

In the main text we discuss chemical shifts predicted using a ML model which extends the ShiftML of Ref.~\cite{paruzzo_2018_shiftml} by training set sparsification and the efficient estimation of the uncertainty in predictions. It is built on the same framework that combines physically-motivated structural representations with a Gaussian process regression (GPR) framework.

\subsection{Sparse Gaussian process regression with uncertainty estimation}

Properties $y$ are predicted from inputs $X$ via an interpolating function $f(X)$ assuming normally distributed noise $\varepsilon \sim \mathcal{N}(0,\sigma)$:
$$ y(X) = \mathnormal{f} (X) + \varepsilon $$
Given a training set of $N$ input-property pairs $({\bf X},{\bf y}) \equiv \left\{\left(X_i\ ,\ y_i\right)\right\}$ one can model $f$ as a Gaussian process $GP\left(0\ ,\ K\right)$, where~ $K$ is the covariance function between the inputs.\\
The prediction for an input $X$ can then be written as a linear combination~\cite{rasmussen_2006_gpr}:
\begin{equation}
    y(X) = \sum_{i=1}^{N} w_i k(X_i,X) = K_{XN} K_{NN}^{-1} {\bf y} \, ,
\end{equation}
where $k(X_i,X) = {(K_{XN})}_i$ and $w_i = \sum_j {(K_{NN}^{-1})}_{ij} y_{j}$.
While predictions can in principle be converged to any desired level of accuracy by including more training data, this rapidly produces kernel matrices $K_{NN}$ of considerable dimensions, slowing down training and predictions. 
We thus follow a projected~process (PP) strategy~\cite{csato_2002_sparsegpr,seeger_2003_sparsegpr,rasmussen_2006_gpr}, in which the full (\(N\times N\)) kernel matrix $K_{NN}$ is approximated by  a lower rank (\(M\times M\)) matrix $K_{MM}$ corresponding to an ``active set'' composed of the $M$ training data which retain the most relevant information. 
The correlations between all the training points and the active set are encoded in an (\(M\times N\)) kernel matrix $K_{MN}$, and predictions for new points $X$ are calculated as
\begin{equation}
    y(X) =  K_{XM}\left( K_{MM} + \varsigma^{-2}K_{MN}K_{MN}^T \right)^{-1} K_{MN}{\bf y}
    \label{eq:pppredictions}
\end{equation}
Here $\varsigma$ is a regularisation parameter.
During training, the size of the matrix to be inverted is thereby reduced to $M\times M$, at the cost of computing, once, the active-passive Gram matrix.
Conversely, when predicting, only similarities between the new structures and the active set have to be considered.

\begin{table}
    \centering
    \begin{tabular}{r|c|c|c|c}
                                        &    H &    C &     N &     O \\ \hline
        cut-off radius $r_c [\ang]$     &  4.5 &  4.0 &   4.5 &   4.5 \\
        Gaussian width $\sigma [\ang]$  &  0.3 &  0.3 &   0.3 &   0.3 \\
        radial basis set size $n$       &   12 &   12 &    12 &    12 \\
        angular basis set size $l$      &    9 &    9 &     9 &     9 \\
        kernel exponent $\zeta$         &    3 &    3 &     3 &     3 \\
        scaling onset $r_s [\ang]$      &  2.0 &  2.0 &   2.0 &   2.0 \\
        scaling exponent $e_s$          &    3 &    3 &     3 &     3 \\
        training set size $N$           &  50k &  50k &   40k &   40k \\
        active set size $M$             &  20k &  20k &   20k &   20k \\
        number of FPS features          & 8000 & 8000 &  8000 &  8000 \\
        regularisation $\varsigma$      & 1800 & 3200 &  5300 &  3000 \\
        test set RMSE  [ppm]            & 0.48 & 4.13 & 13.70 & 17.05
    \end{tabular}
    \caption{SOAP hyperparameters and sparsification parameters for all species.}
    \label{tab:sparsification}
\end{table}

In principle the uncertainty associated with a PP prediction can be calculated directly as
\begin{equation}
\begin{split}
    \sigma(X)^2 = &\varsigma^2 + K_{XX} - K_{XM} K_{MM}^{-1} K_{XM} \\
    &+ K_{XM} (K_{MM} + \varsigma^{-2} K_{NM}^T
K_{NM})^{-1} K_{XM}^{T} .
\end{split}
\end{equation}
This estimate, however, is considerably more demanding than that of $y$. We therefore instead employ the scheme for accurate and efficient uncertainty estimation proposed in Ref.~\cite{musil_2019_uncertainty} which is based on a committee of models.
An ensemble of $N_m$ models is trained on subsamples of the full training set of size $N_s < N$. 
Crucially, the different structural variance covered by the subsamples affects the spread of predictions $\{ y^{(m)}(X) \}$ obtained from the different models $m$. This is corrected for by rescaling 
\begin{equation}
    y^{(m)}(X) \rightarrow {\bar y}(X) + \alpha \left( y^{(m)}(X) - {\bar y}(X) \right)^{\gamma/2 + 1} \; ,
\end{equation}
where ${\bar y}(X) \equiv 1/N_m \sum_m y^{(m)}(X)$, using the constants $\alpha$ and $\gamma$ which maximise the log-likelihood of the rescaled ensemble predictions for a validation set of choice,
\begin{equation}
P(\mathbf{y}|\{ X_n \}_{n=0,1,..} ) = \prod_{n=0}^{N_{\textrm{v}}} \frac{1}{\sqrt{2 \pi \sigma^2(X_n)}}\exp{\frac{(y_n - y(X_n))^2}{2 \sigma^2(X_n)}}
\end{equation}
where $\sigma^2(X) \equiv 1/N_m \textrm{Var}(\{ y^{(m)}(X) \})$ and $N_{\textrm{v}}$ is the size of the validation set. 
In practice we apply a linear rescaling ($\gamma = 0$), for which the log-likelihood is maximised by
\begin{equation}
    \alpha^2 = \frac{1}{N_{\textrm{v}}}\sum_n\frac{(y_n - {\bar y}(X_n))^2}{\sigma^2(X_n)} \; .
\end{equation}

Uncertainties in predictions can then simply be estimated as the standard deviation over the ensemble of models 
\begin{equation}
    \sigma^{\text{ML}}(X) \approx \sqrt{  \frac{\sum_m \left( y^{(m)}(X) - {\bar y}(X) \right)^2}{N_m - 1} } \; .
\end{equation}

It is worth noting that the resultant uncertainties are environment- and model-dependent. Further they are statistical uncertainties which are uncorrelated with the inherent errors of the underlying reference (GIPAW-DFT) data relative to experiment. In consequence they must be added to the GIPAW-DFT error(s) in quadrature.

\begin{figure}[tbhp]
    \centering
    \includegraphics[width=0.42\textwidth]{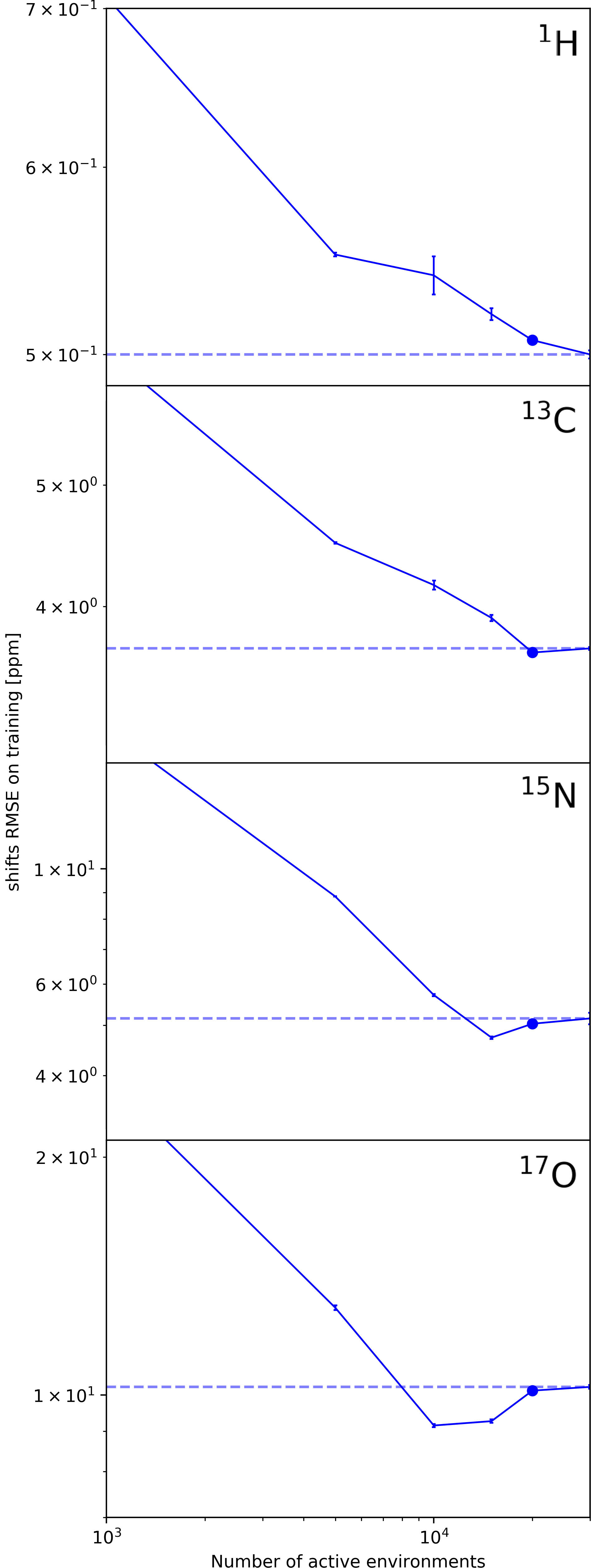}
    \caption{Convergence of the cross-validation RMSE with the number of environments in the active set for $^1$H, $^{13}$C, $^{15}$N, and $^{17}$O.}
    \label{fig:learning_curves_environments}
\end{figure}

\begin{figure}[tbhp]
    \centering
    \includegraphics[width=0.425\textwidth]{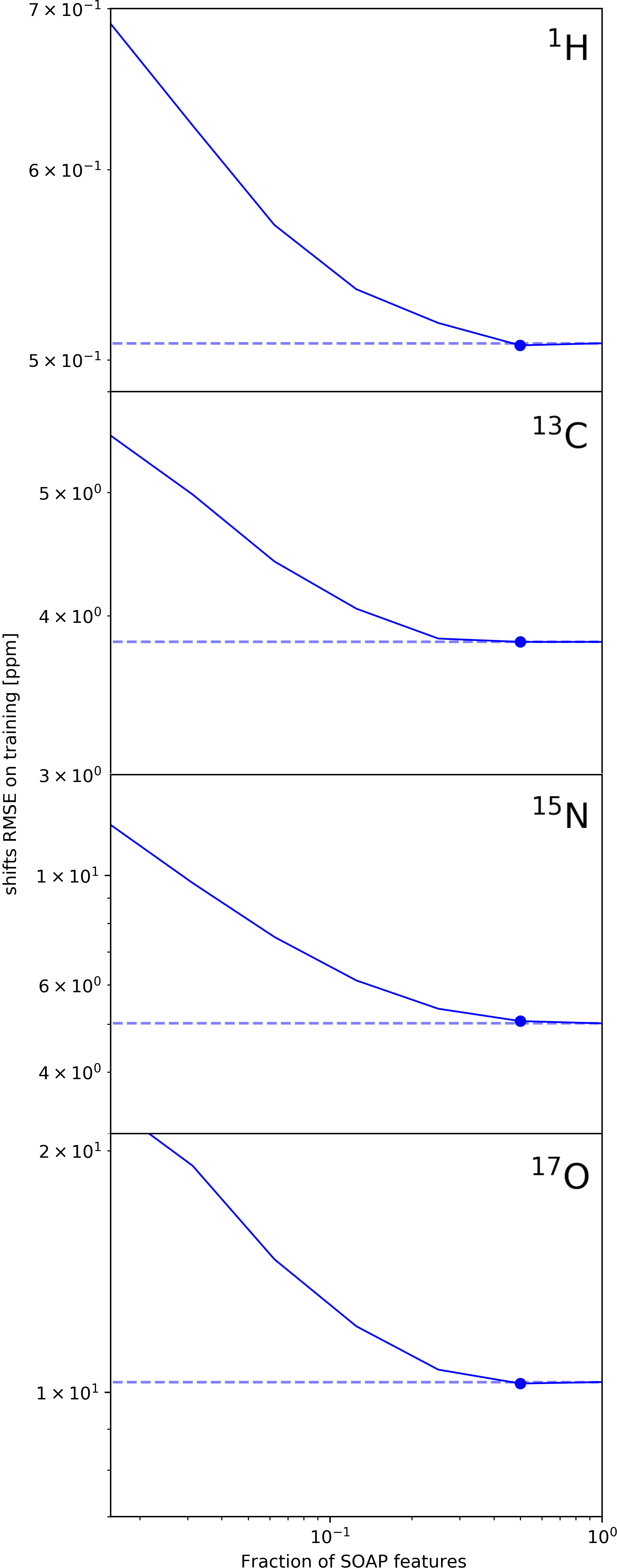}
    \caption{Convergence of the cross-validation RMSE with the fraction of retained SOAP features for $^1$H, $^{13}$C, $^{15}$N, and $^{17}$O.}
    \label{fig:learning_curves_features}
\end{figure}

In practice our GPR model is built around smooth overlap of atomic positions (SOAP) kernels~\cite{bartok_2013_soap,de_2016_soap}, in which atomic environments are represented as species-dependent atomic densities constructed by associating a Gaussian density with each atomic position within a cut-off radius of the central atom.
Using the radially-scaled variant of the SOAP framework~\cite{willatt_2018_rssoap} drastically improves the computational performance compared to the multi-scale approach described in Ref.~\cite{paruzzo_2018_shiftml}, which effectively requires the construction and evaluation of multiple GPR models per chemical species.
The associated hyperparameters were determined using a cross-validation scheme and are detailed in Table~\ref{tab:sparsification}.
The SOAP-GPR framework has proven successful in the context of regressions for different systems~\cite{szlachta_2014,deringer_2017,bartok_2017,musil_2018} and (scalar as well as tensorial) properties~\cite{grisafi_2018_tensorial}. 
Most importantly, SOAP-GPR has previously proven suitable for GIPAW-DFT accurate predictions of NMR chemical shifts~\cite{paruzzo_2018_shiftml}.

\subsection{Training and test set and outlier detection}

A critical element of the ML model are the underlying training and test sets, which were constructed in close analogy to Ref.~\cite{paruzzo_2018_shiftml}. From the around 105,000 crystal structures from the CSD~\cite{groom_2016_csd} which only contain hydrogen, carbon, nitrogen, oxygen, and sulfur and no more than 200 atoms per unit cell, a test set of 604 structures was extracted by random selection and a maximally diverse and informative training set of 3,546 configurations was selected by farthest-point-sampling (FPS)~\cite{eldar_1997_fps,ceriotti_2013_sketchmap,campello_2015_fps} of the remaining structures.
Geometry optimisations and GIPAW NMR calculations were performed using Quantum Espresso~\cite{giannozzi_2009_qe,varini_2013_qe,giannozzi_2017_qe} as detailed in the SI~\cite{supplementary} for all structures in the training and test sets. The training and test sets have been published with Ref.~\citenum{hofstetter_2019_nmr}.

Shifts are calculated for atomic centers, i.e. for local ``environments'', rather than structures. Crystal structures often contain redundant environments, for example due to crystal symmetries. Hence, the training set was reduced in size by FPS ordering the individual environments and retaining only the 100,000 ($^1$H and $^{13}$C) and 40,000 ($^{15}$N and $^{17}$O) most structurally diverse and therefore informative ones.

At this point environments exhibiting GIPAW-DFT shifts far outside the physical ranges of around 5--50\,ppm for $^1$H (64 unphysical environments), around -100--200\,ppm for $^{13}$C (149 unphysical environments), around -700--400\,ppm for $^{15}$N (12 unphysical environments), and around -1250--350\,ppm for $^{17}$O (13 unphysical environments) were eliminated.
Their presence highlights that GIPAW-DFT shifts are not always reliable and hints at the possible presence of further outliers. Initial ML models were therefore trained in a cross-validation scheme to assess (i) the residual error with respect to the GIPAW-DFT reference and (ii) the estimated ML uncertainty for all training environments. 
These were then used to identify anomalous environments with residual errors outside the $3\sigma$ confidence interval associated with the estimated ML uncertainty, suggesting a possible failure of the GIPAW-DFT shift calculation. 
For each anomalous environment the entire associated structure was purged from the training set. 
Since these ML models are to some degree corrupted by the initial presence of anomalous environments, this procedure of training ML models, identifying anomalous environments, and eliminating the associated structures was repeated until the distributions of residual errors were consistent with the estimated ML uncertainties.
We found this procedure to improve the accuracy of the model when applied to the validation set, which suggests that indeed ``outliers'' in the train set affect adversely the accuracy of the model.
All in all 373 $^1$H, 347 $^{13}$C, 44 $^{15}$N, and 113 $^{17}$O environments were eliminated.

\subsection{Active set selction and feature optimization}

Active sets were then extracted on the basis of the FPS order, so as to incorporate the largest amount of information for a given size~\cite{eldar_1997_fps,ceriotti_2013_sketchmap,campello_2015_fps,imbalzano_2018_fps}.
The cross-validation RMSE curves as a function of the size of the active set in Fig.~\ref{fig:learning_curves_environments} suggests that for all species active sets of $M = 20,000$ environments suffice to match the accuracy of the unsparsified models to within less than 1\% of the RMSE of the full model.
It is worth noting that within the PP framework the underlying training set can be arbitrarily large since $K_{MN}K^T_{MN}$ in Eq.~\eqref{eq:pppredictions} can be calculated in chunks, so that the only limiting factor in constructing and applying the PP model is the size of the active set. In practice underlying training sets of $N = 50,000$ for $^1$H and $^{13}$C and $N = 40,000$ for $^{15}$N and $^{17}$O were found to be sufficient to saturate the accuracy of the respective models.

To further accelerate the ML predictions, we also sparsified the SOAP feature vectors, using a FPS strategy~\cite{imbalzano_2018_fps}, performing a separate selection for each element.
Cross-validation demonstrates that the first 20,000 training environments for any given chemical species suffice to guide the FPS of the SOAP features (see SI).
It should be noted that FPS-based choice of SOAP features is guided by structural variability and consequently leads to sparsified fingerprints which should be suitable for regressions of general observables.
The RMSE curves of models built with an increasing number of SOAP features (see Fig.~\ref{fig:learning_curves_features}) demonstrate that sparsifying from an initial 18,301 components to 8000 leads to a negligible decrease in model accuracy for all species (less than 1\% increase in the RMSE).

\begin{figure*}
    \centering
    \begin{tabular}{ll}
        (a) & (b)\\
        \includegraphics[width=0.495\textwidth]{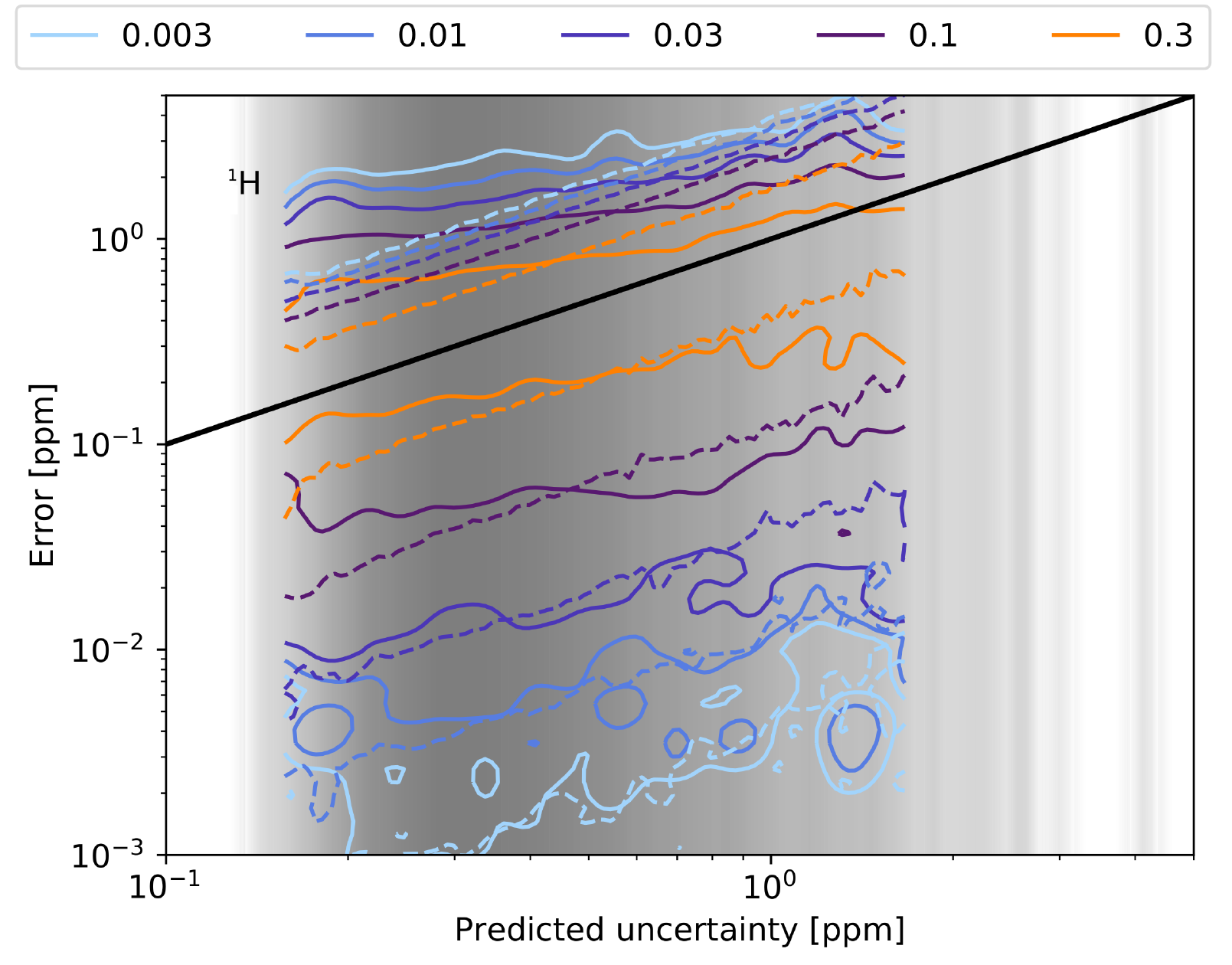} & 
        \includegraphics[width=0.49\textwidth]{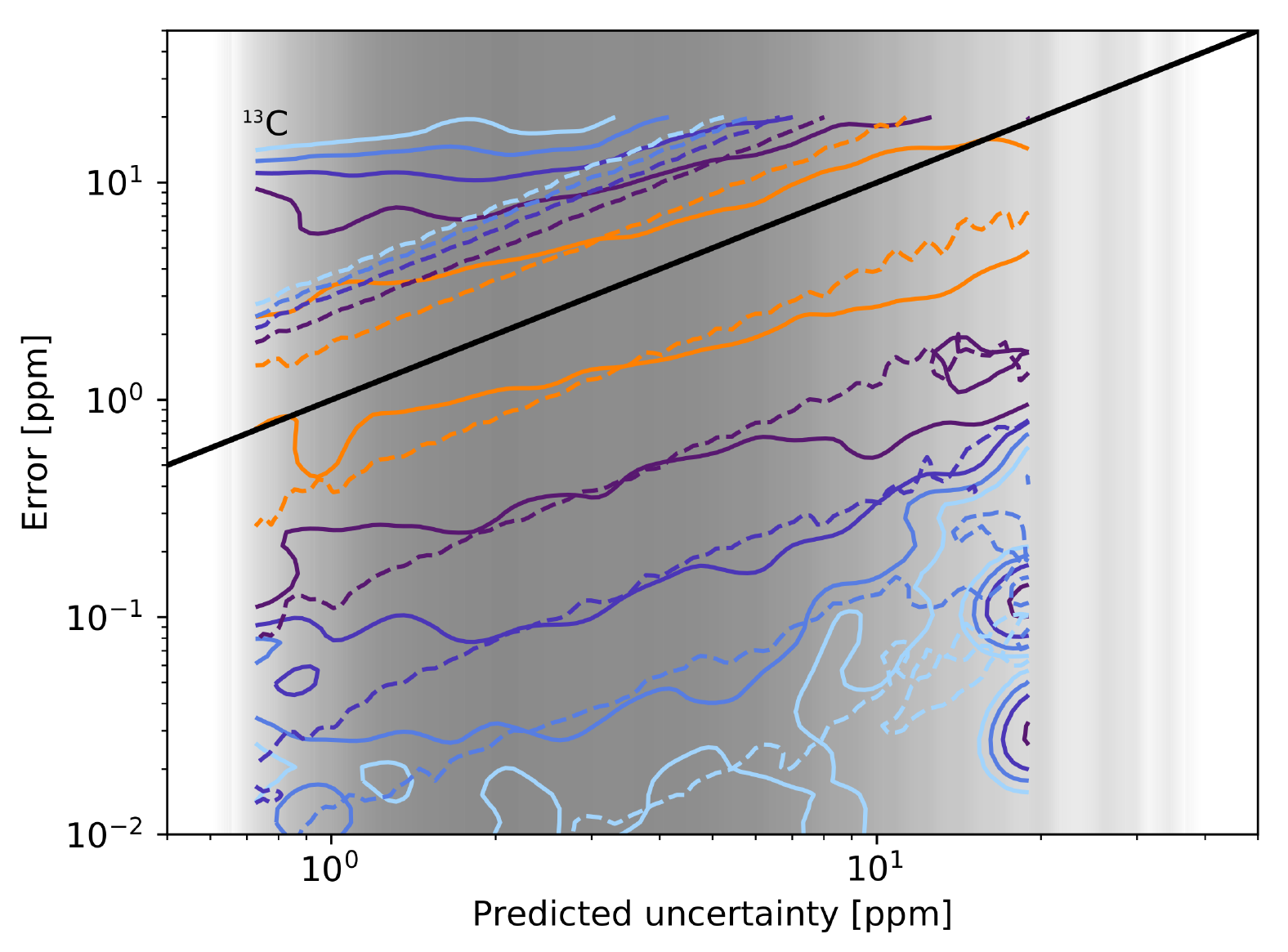}\\
        (c) & (d)\\
        \includegraphics[width=0.49\textwidth]{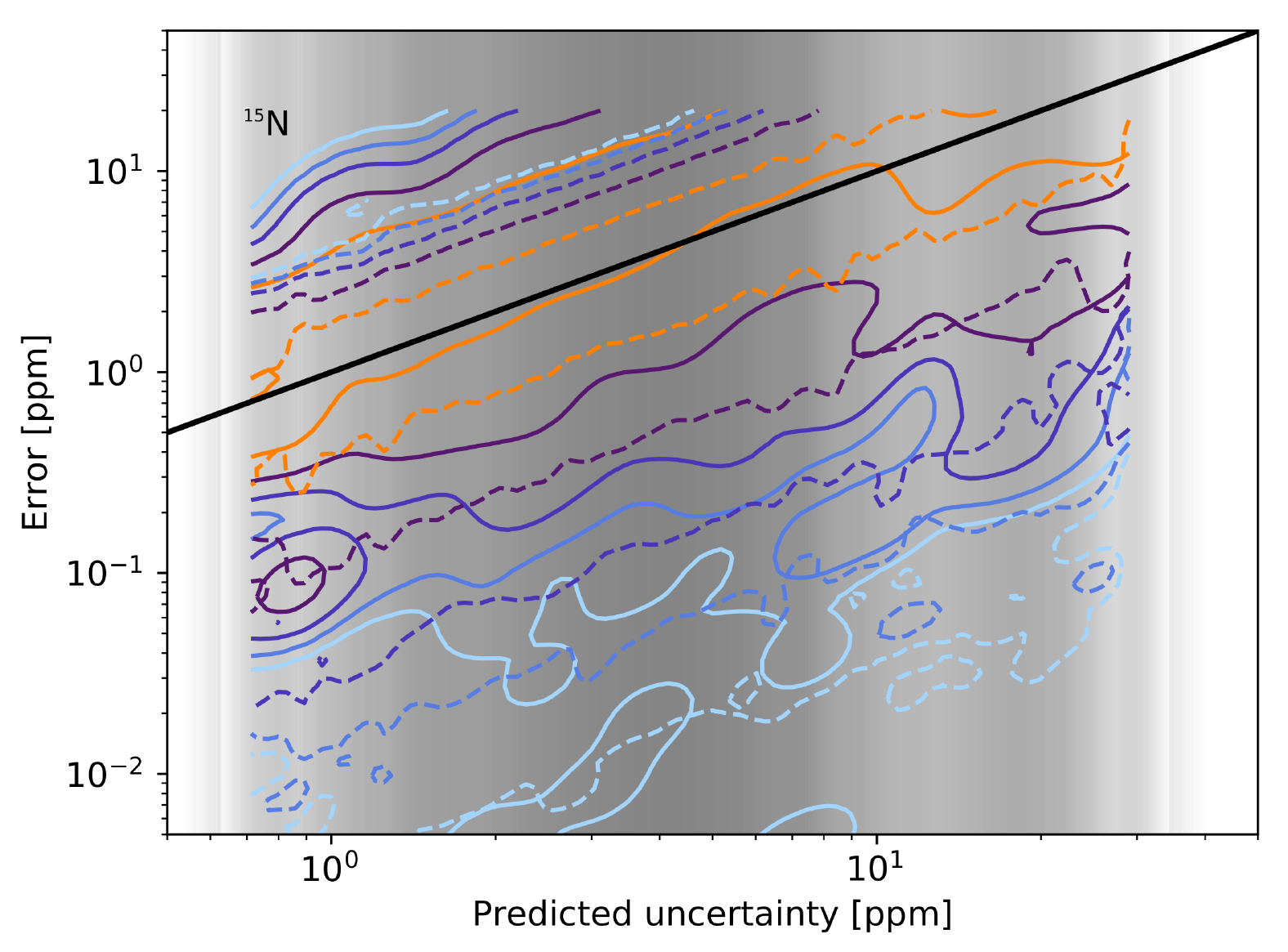} & 
        \includegraphics[width=0.49\textwidth]{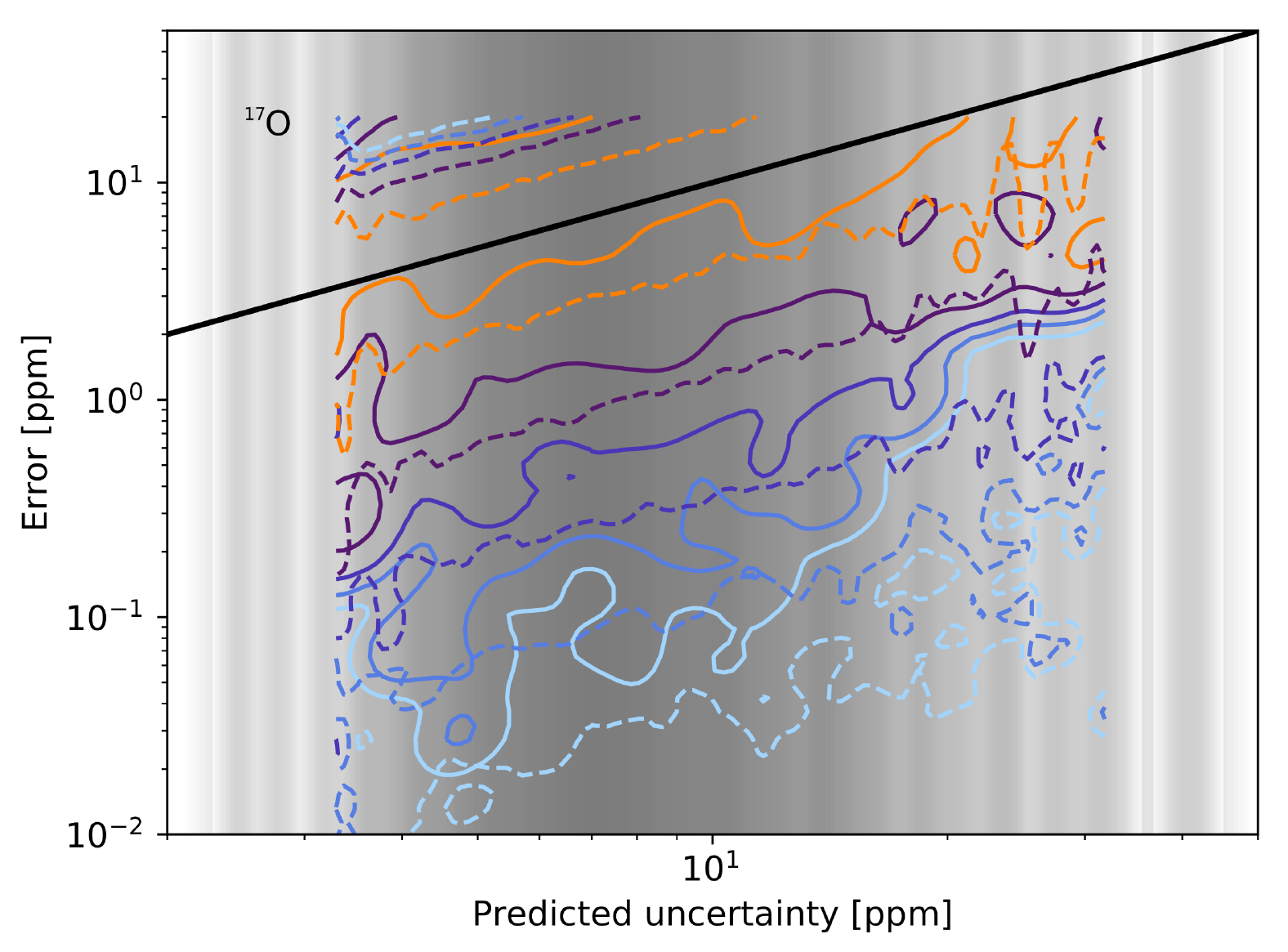}
    \end{tabular}
    \caption{Distribution of (a) $^1$H, (b) $^{13}$C, (c) $^{15}$N, (d) $^{17}$O chemical shielding predictions. The coloured solid lines show contours of the distribution of actual errors relative to the reference, $P(\ln | \overline{y}(X_i) - y_i | | \ln \sigma^{\textrm{ML}}(X))$, while the coloured dashed lines show contours of distribution of the predictions of the subsampling models around their mean, $P(ln | y^{(m)}(X_i) - \overline{y}(X_i) | | \ln \sigma^{\textrm{ML}}(X))$. The gray scale density plot corresponds to the marginal distribution of the predicted uncertainty $P(\ln \sigma^{\textrm{ML}}(X) )$. The solid black line shows $y = x$ to guide the eye.}
    \label{fig:error_likelihood}
\end{figure*}

\subsection{Model performance}

The full sets of hyperparameters defining the specific ML models constructed in this work are collected in Table~\ref{tab:sparsification}.
The final accuracy of this sparse model is (slighlty) better than that of the original ShiftML model presented in Ref.~\cite{paruzzo_2018_shiftml}. 
The error of 0.48\,ppm for out-of-sample predictions of $^1$H shifts on the test set are comparable to the inherent error of the underlying GIPAW-DFT predictions with respect to experiment of around $0.33\pm0.16$\,ppm~\cite{salager_2010_nmr,hartman_2016,dracinsky_2019}.  Further reductions in ML errors would reap insignificant improvements to the resolving power of ML-based NMR crystallography without accompanying reductions in the underlying GIPAW-DFT errors with respect to experiment.
For $^{13}$C the expected ML errors of 4.13\,ppm are about twice as large as the typical error in GIPAW-DFT predictions of $1.9\pm0.4$\,ppm. Even though, as demonstrated in section~\ref{sec:applications}, GIPAW-DFT $^{13}$C errors are often much larger than this value, an improvement in the accuracy of ShiftML for carbon, oxygen and nitrogen would be desirable, and will be the subject of future improvements of ShiftML.

Finally, Fig.~\ref{fig:error_likelihood} demonstrates the agreement between the distributions of ML errors with respect to GIPAW-DFT, $\vert {\bar y}(X_i) - y_i \vert$, and that predicted in terms of the distribution around the mean of the ensemble of subsampling models, $\vert \sum_m y^{(m)}(X_i) - {\bar y}(X_i) \vert$.  The qualitative agreement between the distributions confirms that the standard deviation over the ensemble of models provides a good estimate of the uncertainty in the ML predictions.

\subsection{Model and data availability}

The training and test dataset underlying this version 1.1 of ShiftML, and an on-line tool to evaluate the chemical shifts of molecular crystals containing H, C, N, O, S according to the model, have been made available on the Materials Cloud portal (\url{http://materialscloud.org}) and on \url{http://shiftml.org}.

\section{\label{app:pca}NMR-based similarity kernel}

We construct a matrix of pairwise distances between models (one of which may be experiment) $d(M,M') = -\ln{p(M,M')}$, where $p(M,M')$ is the probability of mistaking $M$ for $M'$ on the basis of shifts measurements.
Momentarily setting aside normalisation, $p(M,M')$ can be calculated as 
\begin{equation}
    \begin{split}
        p(M,M') 
        &= \int \textrm{d}{\bf y} p(M|{\bf y}) p({\bf y}|M') \\
        &= \int \textrm{d}{\bf y} \frac{p({\bf y}|M) p({\bf y}|M')}{p({\bf y}|M) + p({\bf y}|M')}
    \end{split}
    \label{eq:distance_matrix}
\end{equation}
In the limit of infinitesimal uncertainties in the reference shifts, ${\bf y^{M'}}$, this simplifies to 
\begin{equation}
    \lim_{{\bf \sigma}^{M'} \rightarrow \varepsilon} p(M,M') 
    \propto \varepsilon p({\bf y^{M'}}|M)
\end{equation}
which is then symmetrised and normalised, giving 
\begin{equation}
p(M,M') = \frac{p\big( {\bf y^{M'}} \big| M \big) + p\big( {\bf y^M} \big| M' \big)}
{2 \sqrt{p\big( {\bf y^M} \big| M \big) p\big( {\bf y^{M'}} \big| M' \big)}} 
\end{equation}
In the case in which the probability is constructed from fully-assigned shifts, the resulting distance function is proportional  to the squared Euclidean distance between the vectors containing chemical shifts of the various nuclei.
A similarity kernel is then constructed by centering the associated distance matrix $d$
\begin{equation}
    \begin{split}
        k(M,M') &= \sum_{M'',M'''} h(M,M'') d(M'',M''') h(M''',M') \\
        h(M,M') &= \delta_{M,M'} - 1/N_M,
    \end{split}
\end{equation}
and is then used in a KPCA scheme to identify the two principal components on which to represent structural diversity.

\end{document}